\documentclass[traditabstract]{aa}
\usepackage{graphicx}
\usepackage{txfonts}
\usepackage{times}
\begin{document}
\newcommand{\SCARGLE}{{\sffamily\scshape scargle\,}}
\newcommand{\MTRAP}{{\sffamily\scshape mtrap\,}}
\newcommand{\MOST}{{\sffamily\scshape MOST\,}}
\newcommand{\COROT}{{\sffamily\scshape CoRoT}}

\newcommand{\MgII}{\ion{Mg}{ii}}
\newcommand{\HeI}{\ion{He}{i}}
\newcommand{\Halpha}{H$\alpha$}
\newcommand{\Hbeta}{H$\beta$}
\newcommand{\Hgamma}{H$\gamma$}

\def\cd{d$^{-1}$}
\def\cds{d$^{-1}$\,}
\def\kms{km~s$^{-1}$}

\title{Ground-based observations of the $\beta$~Cephei CoRoT main target HD\,180642: abundance analysis and mode identification
\thanks{Based on data gathered with the 1.2m Mercator telescope Roque de los Muchachos, La Palma, the 90cm telescope at Sierra Nevada Observatory, Spain, the 1.5 m telescope at San Pedro M\'{a}rtir Observatory, Mexico, the 1m RCC and 50cm telescope at the Piszk\'estet\H{o} Mountain Station of Konkoly Observatory, Hungary, the 2.2m ESO telescope (ESO Programme 077.D-0311; ESO Large Programme 178.D-0361) at La Silla, Chile, the 1.93m and 1.52m telescopes at the Haute-Provence Observatory, France}}

\author{
M.\,Briquet\inst{1}, 
K.\,Uytterhoeven\inst{2,3,}\thanks{Current address: Laboratoire AIM, CEA/DSM-CNRS-Universit\'e Paris Diderot; CEA, IRFU, SAp, centre de Saclay, F-91191, Gif-sur-Yvette, France},
T.\,Morel\inst{1,4},  
C.\,Aerts\inst{1,5},
P.\,De Cat\inst{6}, 
P.\,Mathias\inst{7},
K.\,Lefever\inst{1,8},   
A.\,Miglio\inst{4}, 
E.\,Poretti\inst{2}, 
S.\,Mart\'{\i}n-Ruiz\inst{9}, 
M.\,Papar\'o\inst{10}, 
M.\,Rainer\inst{2},
F.\,Carrier\inst{1},
J.\,Guti\'errez-Soto\inst{11}, 
J.C.\,Valtier\inst{7}, 
J.M.\,Benk\H{o}\inst{10}, 
Zs.\,Bogn\'ar\inst{10}, 
E.\,Niemczura\inst{1,12},
P.J.\,Amado\inst{9}, 
J.C.\,Su\'arez\inst{9}, 
A.\,Moya\inst{9},
C.\,Rodr\'iguez-L\'opez\inst{9,13,14} 
and R.\,Garrido\inst{9}}

\institute{
Instituut voor Sterrenkunde, Katholieke Universiteit Leuven, Celestijnenlaan 200 D, B-3001 Leuven, Belgium
\and INAF-Osservatorio Astronomico di Brera, Via E. Bianchi 46, I-23807 Merate, Italy 
\and
Instituto de Astrof\'{i}sica de Canarias, Calle Via L\'actea s/n, E-38205 La Laguna, TF, Spain 
\and
Institut d'Astrophysique et de G\'eophysique de l'Universit\'e de Li\`ege, All\'ee du 6 Ao\^ut 17, B-4000 Li\`ege, Belgium 
\and Department of Astrophysics, University of Nijmegen, IMAPP, PO Box 9010, 6500 GL Nijmegen, the Netherlands 
\and
Koninklijke Sterrenwacht van Belgi\"e, Ringlaan 3, B-1180 Brussel, Belgium 
\and
UMR 6525 H. Fizeau, UNS, CNRS, OCA, Campus Valrose, F-06108 Nice Cedex 2, France
\and BIRA-IASB, Ringlaan 3, B-1180 Brussel, Belgium
\and Instituto de Astrof\'{\i}sica de Andaluc\'{\i}a (CSIC), Apdo. 3004, 18080 Granada, Spain 
\and Konkoly Observatory, P.O. Box 67, 1525 Budapest, Hungary 
\and GEPI, Observatoire de Paris, CNRS, Universit\'e Paris Diderot, place Jules Janssen 92195 Meudon Cedex, France
\and Instytut Astronomiczny Wroclawski, Kopernika 11, 51--622 Wroclaw, Poland
\and Laboratoire d'Astrophysique de Toulouse-Tarbes, Universit\'e de Toulouse, CNRS, Toulouse 31400 France
\and Universidade de Vigo, Departamento de F\'isica Aplicada, Campus Lagoas-Marcosende, Vigo 36310 Spain}

\authorrunning{Briquet et al.} \titlerunning{Ground-based observations of the $\beta$~Cephei CoRoT main target HD\,180642} \date{Received ; accepted}

\abstract{The known $\beta$\,Cephei star HD\,180642 was observed by the CoRoT satellite in 2007. From the very high-precision light curve, its pulsation frequency spectrum could be derived for the first time (Degroote and collaborators). In this paper, we obtain additional constraints for forthcoming asteroseismic modeling of the target. Our results are based on both extensive ground-based multicolour photometry and high-resolution spectroscopy. We determine $T_{\rm eff} =$ 24\,500$\pm$1000 K and $\log g =$ 3.45$\pm$0.15 dex from spectroscopy. The derived chemical abundances are consistent with those for B stars in the solar neighbourhood, except for a mild nitrogen excess. A metallicity $Z =$ 0.0099$\pm$0.0016 is obtained. Three modes are detected in photometry. The degree $\ell$ is unambiguously identified for two of them: $\ell = 0$ and $\ell = 3$ for the frequencies 5.48694\,d$^{-1}$ and 0.30818\,d$^{-1}$, respectively. The radial mode is non-linear and highly dominant with an amplitude in the U-filter about 15 times larger than the strongest of the other modes. For the third frequency of $7.36673$\,d$^{-1}$ found in photometry, two possibilities remain: $\ell = 0$ or 3. In the radial velocities, the dominant radial mode presents a so-called stillstand but no clear evidence of the existence of shocks is observed. Four low-amplitude modes are found in spectroscopy and one of them, with frequency 8.4079\,d$^{-1}$, is identified as $(\ell,m)=(3,2)$. Based on this mode identification, we finally deduce an equatorial rotational velocity of 38$\pm$15 km s$^{-1}$.}{}{}{}{}


\maketitle

\section{Introduction}

\begin{table*}
\caption{Logbook of the photometric observations of HD\,180642. }
\label{logbook1}
\begin{center}
\begin{tabular}{llrllll} \hline  \hline
Observatory & N & $\Delta T$ & HJD begin & HJD end & filter(s) &$f_1$ \\
\hline
La Silla & 20 & 186.7 & 50562.8 & 50749.6 & UB$_1$BB$_2$V$_1$VG & 5.4870(2) \\
La Palma & 171 & 1183.8 & 52421.6 & 53605.4 & UB$_1$BB$_2$V$_1$VG & 5.48693(1) \\
SPMO & 113 & 5.3 & 53192.7 & 53198.0 & $u,v,b,y$ & 5.494(3) \\
SNO & 222 & 359.2 & 53205.4 &  53564.6 & $u,v,b,y$ & 5.48693(3)\\
KO  & 84 & 319.2 & 53260.3 & 53579.5 & Johnson $V$ & 5.49(1)\\ 
KO CCD & 261 & 26.1 & 53921.4 & 53947.5 & Johnson $V$ & 5.4858(3)\\
ASAS & 363 & 2797.6 & 51979.9 & 54777.5 & Johnson $V$ & 5.486915(7)\\
\hline
Str\"omgren & 335 & 371.9 &  53192.7 & 53564.6 & $u,v,b,y$ & 5.48693(2) \\
U & 526 & 3042.6 & 50562.8 & 53605.4 & U, $u$ & 5.48695(2) \\
B & 526 & 3042.6 & 50562.8 & 53605.4 & B, $v$ & 5.486947(2) \\
V & 526 & 3042.6 & 50562.8 & 53605.4 & V, $y$ & 5.486944(2) \\
V+ & 1234 & 4214.7 & 50562.8  & 54777.5 & V, $y$, $V$ & 5.486938(5) \\
\hline
\end{tabular}
\end{center}
\tiny{The number N of datapoints retained for the analysis, the total timespan $\Delta T$ (in days) and the available filters are given. The last column gives the value of the dominant frequency (see Sect.~\ref{freqsection}) expressed in d$^{-1}$ with its error between brackets. The HJD is given with respect to $HJD_0=$ 2400000. Str\"omgren refers to merged SPMO and SNO Str\"omgren data. The merged Geneva U and Str\"omgren $u$, B and $v$, and V and $y$ data, are denoted as U, B, V data, respectively. The V+ dataset corresponds to the V data to which the KO Johnson $V$ data and ASAS Johnson $V$ data are also added.} 
\end{table*}

The $\beta$~Cephei stars are a homogeneous group of oscillating B0--B3 stars that have been studied as a class for more than a century. Stankov \& Handler\ (\cite{stankov_handler}) compiled an overview of the observational properties of this group of stars. The oscillations of $\beta\,$Cephei stars are explained in terms of the $\kappa$\,mechanism that operates on iron-peak elements (Dziembowski \& Pamyatnykh\ \cite{dziembowski_pamyatnykh}; Gautschy \& Saio\ \cite{gautschy_saio}). Given that mainly low-degree low-order pressure and gravity modes are observed, these stars are good potential targets for in-depth asteroseismic studies of the interior structure of massive stars. 

In principle, these studies would allow us to calibrate the evolutionary models across the entire main sequence, although most $\beta$\,Cephei stars seem to be observed while in the second part of their main-sequence lifetime (Stankov \& Handler\,\cite{stankov_handler}). Moreover, theory predicts the occurrence of stellar oscillations in B stars after the main sequence (Saio et al.\,\cite{saio}) and oscillations in such stars have indeed been found (Lefever et al.\,\cite{lefever07}).

Asteroseismic studies have been accomplished for 5 $\beta\,$Cephei stars only (see Aerts\ \cite{aerts08} for a review). These studies were based on intensive ground-based observing campaigns and led, for the first time, to constraints on the core overshoot and internal rotation of massive B-type stars. Despite these successes, we clearly need to increase the number of asteroseismically determined models for these types of stars. This will become especially important in the near future to fully exploit the data becoming available from ongoing dedicated ground-based observing campaigns (see Uytterhoeven\ \cite{uytterhoeven09} for a review) as well as ongoing space missions such as MOST, CoRoT and Kepler.

The B\,1.5\,II-III star \object{HD\,180642} (V1449 Aql, HIP\,94793, \mbox{V = 8.29 mag}) was discovered as a candidate $\beta\,$Cephei star during the Hipparcos mission by Waelkens et al.\ (\cite{waelkens}). Afterwards, Aerts\ (\cite{aerts00}) used Geneva photometry to identify its dominant mode as a radial one, confirming it to be a $\beta\,$Cephei star. Moreover, HD\,180642 is among known $\beta\,$Cephei stars of the highest amplitude. Its peak-to-peak V amplitude in the Geneva filter is 0.078 mag. For comparison, the mean value for the $\beta$~Cephei stars in the catalog of Stankov \& Handler\ (\cite{stankov_handler}) is $\sim$0.025 mag, for visual filters. 

Our studied target was observed by the CoRoT satellite during a long run (156 days) between May and October 2007. We note that it is the only known $\beta$~Cephei star in CoRoT's core programme. A detailed modeling and interpretation of the CoRoT data is presented in Degroote et al.\ (\cite{degroote}). These latter authors modelled this very high-precision light curve with 11 independent frequencies and 22 low-order sum and difference combination frequencies. Nine of the independent frequencies are in the range expected for $\beta$~Cephei stars, between 5 and 9\,d$^{-1}$. Moreover, 5 of the 33 frequencies are in the regime of high-order \mbox{g modes} with frequencies below 2\,d$^{-1}$. 

To complement the space white-light data, we conducted a project to collect ground-based multicolour photometry and high-resolution spectroscopy for selected primary and secondary CoRoT targets, and we refer to Poretti et al.\ (\cite{poretti}) and Uytterhoeven et al.\ (\cite{uytterhoeven08}) for a description of this campaign. In this paper, we present the results of the analysis of the ground-based datasets for HD\,180642. We aim to determine the fundamental parameters and chemical abundances of the object and to derive the wavenumbers $(\ell,m)$ of its low-amplitude modes, which are also detected in our ground-based measurements. 
 
The paper is divided into two parts. Sections\,2 and 3 are devoted to photometry and spectroscopy, respectively. In both sections, we first describe our observations and data reduction. Afterwards, we describe our frequency analysis followed by our mode identification. In the spectroscopic part, we also present an abundance analysis. We end the paper with a summary of our results in Sect.\,4. 

\section{Photometry}

\begin{figure*}
\centering
\rotatebox{-90}{\resizebox{5.cm}{!}{\includegraphics{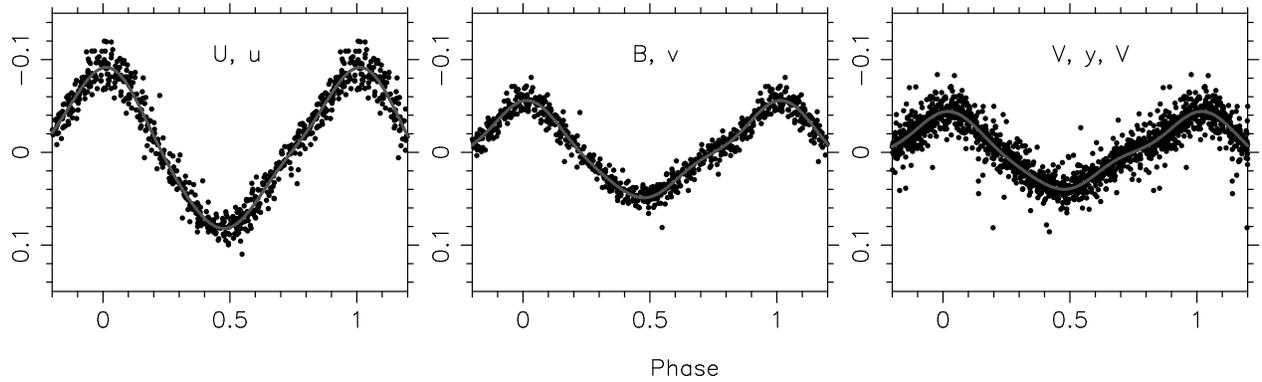}}}
\caption{Phase diagrams for the ultraviolet, blue, and visual merged datasets  U, B, V+ (dots) of HD\,180642 folded according to the dominant frequency $f_1=5.48694$\,d$^{-1}$. The full gray lines are leastsquares fits including $f_1$, $2f_1$ and $3f_1$.}
\label{fases1}
\end{figure*}

\subsection{Data description}

The star was monitored with the 1.2m Mercator telescope (MER) at La Palma in the framework of a long-term photometric monitoring programme dedicated to pulsating stars. For our target, we completed observing campaings of Geneva seven-colour (UB$_1$BB$_2$V$_1$VG) high-precision photometry during two seasons (2002, 2005) with 80-day and 57-day durations, respectively. The star was observed a few times every clear night. In total, we used 171 \mbox{7-colour} Geneva measurements with a time span of 1184 days. Aerts\ (\cite{aerts00}) had already obtained 20 datapoints of the star with the same instrument attached to the 0.7m Swiss telescope at La Silla. 

With the goal of gathering additional measurements with a higher sampling per night, HD\,180642 was observed with the twin Danish six-channel {\sl uvby}$\beta$ photometers attached to the 90cm telescope at Sierra Nevada Observatory (SNO), Spain, and at the 1.5\,m telescope at San Pedro M\'{a}rtir Observatory (SPMO), Mexico. All the data were collected in the four Str\"omgren {\sl uvby} filters. At SNO, a total of 222 high quality datapoints were collected during 1 night in July 2004, 5 nights in June 2005, and 6 clear nights in July 2005. At SPMO, our target was observed during 4 nights in July 2004, resulting in a total of 113 data points.

Johnson $V$ measurements were also obtained with the 50\,cm telescope at the Piszk\'estet\H{o} Mountain Station of Konkoly Observatory (KO) during 2 nights in September 2004 and 3 nights in July 2005. Since the $V$ data showed many scattered points, we removed all data deviating by more than 0.05 mag from the fit with the dominant frequency (see Sect.~\ref{freqsection}). We retained a total of 84 datapoints for the analysis. The comparison star used at all sites was HD\,181414 (\mbox{V = 7.068 mag}, A2). 

During a total of 5 nights in July and August 2006, HD\,180642 was also observed in Johnson $V$ filter with a Princeton Instruments VersArray 1300B CCD detector attached to the 1m RCC telescope (f/13.5) at KO. The data were binned such that the total integration time of the binned datapoints was about 3 minutes. In total, we retained 261 datapoints. Since the same comparison stars were monitored not every night in the CCD frame, we reconstructed  relative magnitude differences using common bright targets.

Finally, we used the Johnson $V$ ASAS publically available photometry from Pigulski \& Pojma\'nski\ (\cite{pigulski_pojmanski}). We do not include the Hipparcos photomety in our analysis because the frequency value of the dominant mode (5.48709(3) d$^{-1}$) in these earlier observations, differs significantly from the values derived in the other photometric measurements (see Sect.\,\ref{freqsection}). A summary of the datasets used is listed in Table\,\ref{logbook1}.

\subsection{Frequency analysis}\label{freqsection}

\begin{figure}
\centering
\rotatebox{-90}{\resizebox{7.2cm}{!}{\includegraphics{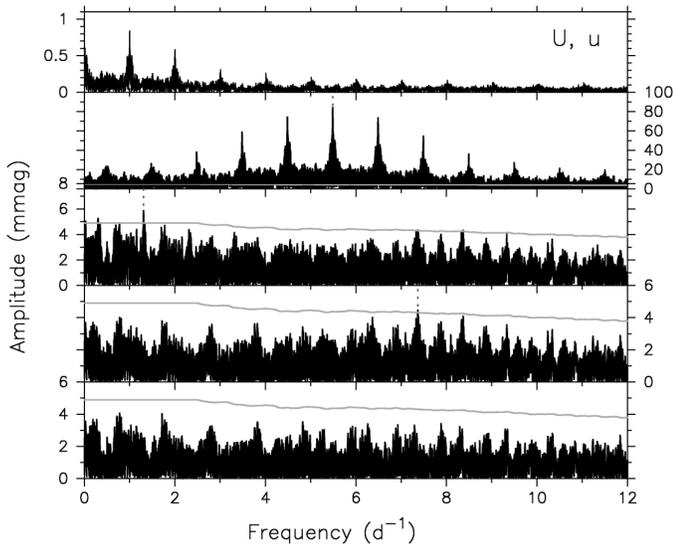}}}
\caption{\SCARGLE periodograms calculated from the combined Geneva $U$ and Str\"omgren $u$ data. The uppermost panel shows the spectral window of the data. All subsequent panels show the periodograms at different stages of prewhitening. From top to bottom: periodogram of the observed light variations, of the data prewhitened with $f_1$ and its two harmonics $2f_1$ and $3f_1$, of the data subsequently prewhitened with $f_2^p$, and of the data subsequently prewhitened with $f_3^p$. The three frequencies found are indicated by dashed gray lines. The light gray lines indicate the 4 signal-to-noise (S/N) level. It is computed as the average amplitude over a frequency interval with a width of 2~\cds in an oversampled \SCARGLE periodogram obtained after final prewhitening. The highest peak in the third panel corresponds to a one-day alias (1+$f_2^p$) of the true frequency $f_2^p$ (see text).
}
\label{freqs}
\end{figure}

\renewcommand{\arraystretch}{1.2}
\begin{table*}
\caption{Amplitudes and phases of the leastsquares fit to the merged U, B and V+ light curves of HD\,180642.}
\label{ampl}
\begin{center}
\begin{tabular} {l cc cc cc}
\hline
 \hline
Freq& \multicolumn{2}{c}{U \& $u$} & \multicolumn{2}{c}{B \& $v$} & \multicolumn{2}{c}{V \& $y$ \& $V$}  \\ \hline
  (\cd)  & Ampl & Phase & Ampl & Phase & Ampl & Phase    \\ 
    &(mmag)& (rad)&(mmag)& (rad)&(mmag)& (rad) \\ \hline
$f_1 = 5.48694$ & 82.7(7) & 2.884(9) & 48.1(5) & 2.88(1) & 38.1(5)   &  2.86(1) \\
$2f_1 = 10.97388$ & 8.2(7) & 1.48(9) & 6.6(5) & 1.52(8) & 5.3(5) & 1.4(1)   \\
$3f_1 = 16.46082$ & 4.3(7) & 2.0(2) & 4.3(5) & 2.1(1) & 3.6(5) & 1.9(2)   \\
$f_2^p = 0.30818$ & 4.9(7) & 0.9(2) & 3.1(5) & 1.1(2) & 1.6(5) & 0.9(4)  \\
$f_3^p = 7.36673$ & 4.5(7) & 5.4(2) & 3.1(5) & 5.4(2)  & 2.2(5)  &  5.7(3)  \\
\noalign{\smallskip}
\hline
Var. Red. & \multicolumn{2}{c}{96.6\%}  & \multicolumn{2}{c}{95.3\%}& \multicolumn{2}{c}{81.8\%}\\
Residual rms (mmag) & \multicolumn{2}{c}{11.4}  & \multicolumn{2}{c}{7.9}& \multicolumn{2}{c}{12.9} \\
\noalign{\smallskip}
\hline
\end{tabular}
\end{center}
\tiny{Errors in units of the last digit are given in brackets. The amplitude is expressed in millimag, and the phase is given in radians. The adopted reference epoch for $\phi=0.0$ corresponds to HJD 2447963.2564. For each combined dataset the total variance reduction of the quintu-periodic model and the root mean square (rms) of the residuals is reported.} 
\end{table*}

We first searched for frequencies in all filters of the individual datasets, using the \SCARGLE (Scargle\ \cite{scargle}) method and the leastsquares power spectrum method (Van\'{\i}\v{c}ek\ \cite{vanicek}). These methods produced identical results within their respective uncertainties, so we list only the outcome obtained for the \SCARGLE periodograms. In all datasets and filters, we confirm the dominant frequency. The value of $f_1$ found in each separate dataset is given in the last column of Table\,\ref{logbook1}, where the uncertainty in the last digit is given in parenthese. This frequency uncertainty was calculated to be $\sigma_{f} = \sqrt(6) \sigma_{{\rm std}} / \pi \sqrt(N) A_{f} \Delta T$ (Montgomery \& O'Donoghue\ \cite{montgomery_donoghue}), where $\sigma_{{\rm std}}$ is the standard deviation of the final residuals, $A_{f}$ the amplitude of the frequency $f$, and $\Delta T$ the total timespan of the observations. 

To increase the frequency precision, we combined datasets for given filters whenever possible. We merged the SPMO and SNO Str\"omgren data (termed Str\"omgren hereafter), and constructed ultraviolet, blue, and visual ground-based light curves (U,B,V hereafter) by merging the Geneva U and Str\"omgren $u$, B and $v$, and V and $y$ data, respectively. To combine the data, we calculated the preliminary frequency solution of the individual subsets, checked that the amplitudes of the dominant mode in the Geneva/Str\"omgren U/$u$, B/$v$ and V/$y$ are the same within the errors, and realigned the subsets at the same mean brightness level. For a description of the procedure, we refer to Uytterhoeven et al.\ (\cite{uytterhoeven08}, their Sect.\,2.1). We note that in the SNO dataset it was also necessary to align the data obtained in 2004 with the data taken in 2005. 

We also constructed a more extensive visual band light curve (termed V+ hereafter) by also adding the KO Johnson $V$ data and the ASAS Johnson $V$ data to the Geneva V and Str\"omgren $y$ data. The results for the dominant frequency of these merged sets are also listed in Table\,\ref{logbook1}. The refined value of the dominant frequency for each of U, B, V, and V+ is $f_1=5.48694\,$d$^{-1}$.

A harmonic fit to the merged U, B, V light curves is shown in Fig.\,\ref{fases1} and infers the prominent presence of harmonics of $f_1$. The harmonics of $f_1$ were also derived by a frequency analysis of the residual light curves after prewhitening with $f_1$. We found that $4f_1$ no longer has a significant amplitude.

In the periodograms of the residuals of the individual and merged datasets, the highest power occurs at low frequency between 5 and 9\,d$^{-1}$. In the individual datasets, none of the candidate frequencies reaches a high enough amplitude to be significant, where we adopt the criterion derived empirically by Breger et al.\ (\cite{breger}) and studied further by Kuschnig et al.\ (\cite{kuschnig}), who studied the variability of HST guide-stars. They concluded that a peak reaching at least 4 times the noise level in the periodograms has only a 0.1\% chance of having been produced by noise. These results were confirmed by De Cat \& Cuypers\ (\cite{decat_cuypers}) for the case of g-mode pulsations studied with single-site data. The noise level in Fourier space must be computed at each prewhitening stage, after checking how the noise level had decreased because of prewhitening. 

In the combined Str\"omgren and ultraviolet datasets, however, we found evidence of two additional significant frequencies. In the Str\"omgren dataset, we detect 0.2965$\pm1$\,d$^{-1}$ and 7.3697\,d$^{-1}$, while the ultraviolet data infer the frequencies 1.30818$\pm1$\,d$^{-1}$ and 7.36673\,d$^{-1}$. Similar frequencies are also present in the blue and visual datasets, but are of low amplitude. The results from the spectroscopic analysis (see Sect.\,\ref{spectro_FA}) enable us to identify the correct alias peak, and we accept the following additonal frequencies from the photometric data of HD\,180642: $f_2^p=$0.30818(5)\,d$^{-1}$ and $f_3^p=$7.36673(7)\,d$^{-1}$. Figure\,\ref{freqs} shows the \SCARGLE periodograms calculated from the combined Geneva U and Str\"omgren $u$ data (ultraviolet). The residual data (bottom part Fig.~\ref{freqs}) still show signs of frequencies at low amplitude, but to detect them we would need high-accuracy data of longer time span. The U, B, and V datasets, after prewhitening with $f_1, 2f_1$, and $3f_1$, folded with $f_2^p$ and $f_3^p$ are shown in Fig.~\ref{fases1}. The amplitudes and phases for $f_1, 2f_1, 3f_1, f_2^p$, and $f_3^p$ of the leastsquares fit to the merged U, B, and V+ light curves of HD\,180642 are listed in Table\,\ref{ampl}. 

Within their estimated errors, neither $f_2^p$ nor $f_3^p$ is detected in the CoRoT light curve. The frequency values found in the space photometry are \mbox{$f_2^{c_1} = $ 0.29917(9)\,d$^{-1}$} and \mbox{$f_3^{c_1} = $ 7.3586(2)\,d$^{-1}$} (Degroote et al.\ \cite{degroote}). In addition, after prewhitening the main peaks, the frequencies \mbox{$f_2^{c_2} = 0.3204(3)$ \,d$^{-1}$} and \mbox{$f_3^{c_2} =$ 7.3744(3)\,d$^{-1}$} are significant in the CoRoT periodogram. We point out that \mbox{$f_2^p \sim (f_2^{c_1}+ f_2^{c_2})/2$} and \mbox{$f_3^p \sim (f_3^{c_1}+ f_3^{c_2})/2$}. Both sets of three close frequencies cannot be interpreted as rotational multiplets since it would require a configuration such that the central peak is seen in the ground-based data and not in the space data. We also exclude the possibility of them being independent frequencies coincidentally close to each other with a largely varying beat phenomenon because it is unlikely to happen twice. Another possiblity is that these close frequencies are attributed to one and the same mode. This interpretation is only plausible if both the error estimates of frequencies are largely underestimated and the assumption of constant amplitude and phase during prewhitening of $f_2^{c_1}$ and $f_3^{c_1}$ from the CoRoT light curve produces an artificial additional frequency peak close by. The analysis of the very high-precision CoRoT light curve indeed showed that the amplitudes, the phases, and even the frequencies themselves, $f_2^{c_1}$ and $f_3^{c_1}$, are not necessarily constant in time (Degroote et al.\ \cite{degroote}), while constant phases and amplitudes for each frequency are traditionally assumed to estimate the values of these quantities and their errors as well as to prewhiten. It is unclear if this variability in the amplitudes, phases, and/or frequencies could be linked to the changing position of the satellite with respect to the Sun in its 5-month orbit or modes of truly time-variable behaviour. In any case, our ground-based photometry is not accurate enough to observe any amplitude and phase variations with time. Its time base (3043 days) is much longer than the time base of the CoRoT observations (156 days). We consequently interpret the two triple frequency sets each being caused by a single oscillation mode, and we keep the ground-based frequency values to perform the mode identification from these data. 

\subsection{Mode identification}\label{MIsection}

\begin{figure*}
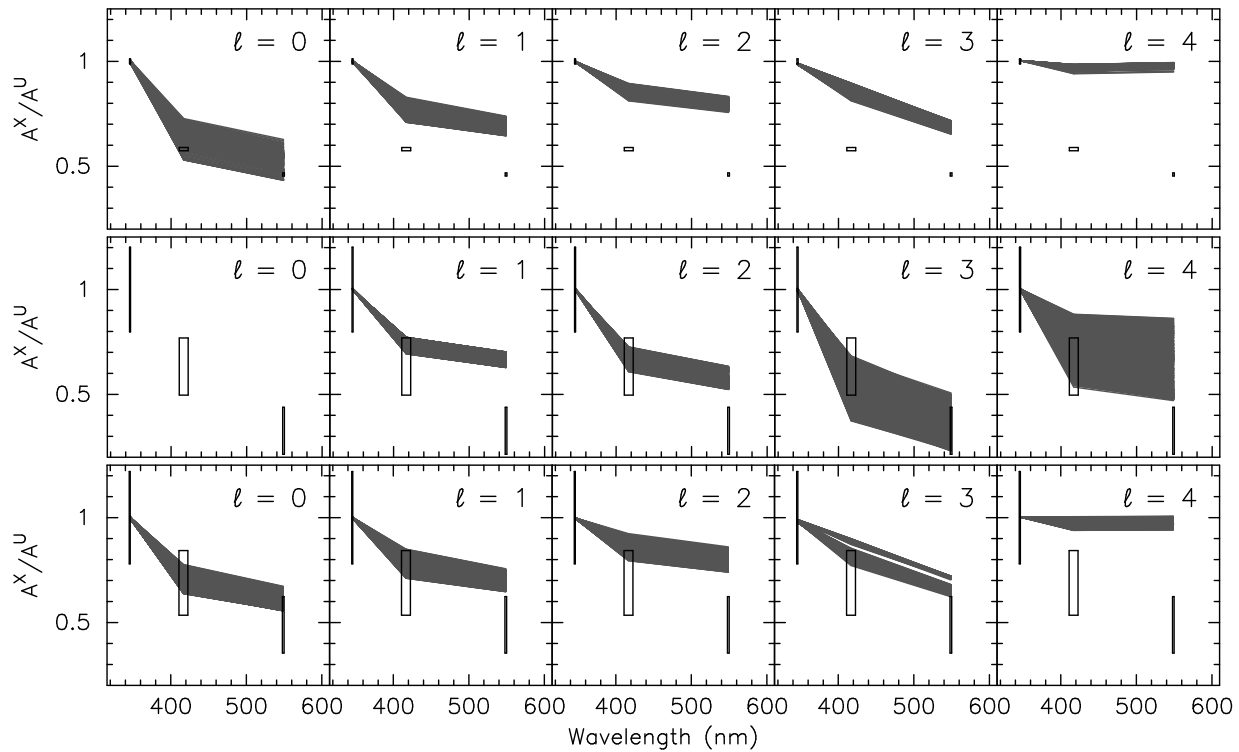

\centering
\hspace{0cm}\rotatebox{-90}{\resizebox{3cm}{!}{\includegraphics{hd180642_photmodeID_mode1_with_0H1H_accepted.ps}}}

\hspace{0cm}\rotatebox{-90}{\resizebox{3cm}{!}{\includegraphics{hd180642_photmodeID_mode2_with_mary_0H1H_accepted.ps}}}

\hspace{0.15cm}\rotatebox{-90}{\resizebox{3.835cm}{!}{\includegraphics{hd180642_photmodeID_mode3_with_mary_0H1H_accepted.ps}}}
\caption{Results of the photometric mode identification for $f_1$ (top), $f_2^p$ (middle) and $f_3^p$ (bottom). The vertical lines indicate the observed amplitude ratios computed from the merged U, B, V+ datasets (see Table\,\ref{ampl}), with their standard error. The theoretical amplitude ratios are represented for modes of different $\ell$-values.
}
\label{ratios}
\end{figure*}

\begin{table*}
\caption[]{Logbook of the spectroscopic observations of HD\,180642 obtained between June 2005 and July 2007 (HJD 2453536.4--2454302.8; $\Delta T = 967.5$ days).}
\label{logbook2}
\begin{center}
\begin{tabular}{cccccccccc}
\hline\hline
 instrument & N & HJD begin & HJD end &$\Delta$T & $<$S/N$>$ & S/N-range & T$_{\rm exp}$ & resolution & wavelength range\\ 
\hline
FEROS & 224 & 2453881 & 2454302  & $967.5$ & 186 & $[100,300]$ & $[300,1200]$ & 48\,000 & $[3570,9215]$\\
SOPHIE & 24 & 2454265 & 2454275 & $10.1$ & 106 & $[80,140]$ & $[1400,1500]$ & 40\,000 & $[3873,6945]$\\
Aur\'elie & 14 & 2453536 & 2453544 & $8.1$ & 93 & $[80,105]$ & 900 & 25\,000 & $[4530,4673]$\\
\hline
\end{tabular}
\end{center}
\tiny{For each instrument the number of high-quality spectra N (i.e., spectra with a S/N-ratio $>$ 80 calculated in the region near 4550\ $\AA$), the HJD, the time-span (in days), the average S/N-ratio, the range of S/N-ratios, the typical exposure times (in seconds), the resolution of the spectrograph and the wavelength range (in $\AA$) are given.} 
\end{table*}

To identify the modes, we applied the usual method of photometric amplitude ratios, using the formalism of Dupret et al.\ (\cite{dupret03}). This technique has already been successfully applied to the pressure modes in $\beta$~Cephei stars (e.g. Aerts et al.\ \cite{aerts06}) and to the gravity modes in slowly pulsating B stars (De Cat et al.\ \cite{decat}).

We computed the non-adiabatic eigenfunctions and eigenfrequencies required for modes with $\ell \le 4$ by using the code called MAD (Dupret et al.\ \cite{dupret01}). Ground-based data are currently not precise enough to detect higher degree modes since cancellation effects leave a very tight signal that is below the detection threshold. The non-adiabatic computations were performed on two grids of theoretical models for which the adopted input physics were different, as described below. All the models were computed with the Code Li\'egeois d'\'Evolution Stellaire (CL\'ES, Scuflaire et al.\ \cite{scuflaire}). 

The first grid (grid\,1 hereafter) is adopted from De Cat et al.\ (\cite{decat}) and we refer to this paper for details of the input physics. The second grid (grid\,2 hereafter) was computed using different opacity tables and different atmosphere models. For grid\,2, OP opacity tables (Seaton\ \cite{seaton}), assuming Asplund et al.\ (\cite{asplund}) metal mixtures, were chosen instead of OPAL opacity tables (Iglesias \& Rogers\ \cite{iglesias_rogers}). Grey atmosphere models were used instead of Kurucz atmosphere models. No overshooting was assumed in grid\,1 whereas different values of the overshooting parameter (between 0 and 0.5 local pressure scale heights) were considered in grid\,2. Finally, grid\,2 covers higher masses (up to 18 M$_\odot$) than grid\,1 (up to 15 M$_\odot$).

The computation of grid\,2 was motivated by the following reasons. Since the paper by De Cat et al.\ (\cite{decat}), work on the modeling of $\beta$~Cephei stars pointed out that OP opacity tables can account for the excitation of modes observed in studied objects, while OPAL opacity tables fail (Miglio et al.\ \cite{miglio}, Briquet et al.\ \cite{briquet07}, Dziembowski \& Pamyatnykh\ \cite{dziembowski_pamyatnykh08}). In addition, closer agreement between theory and observations is also achieved if overshooting is present (Briquet et al.\ \cite{briquet07}, Aerts\ \cite{aerts08}). 

We obtained the same conclusions regardless of the grid considered. In the following, we present our mode identification outcome for grid\,2. We used the models whose stellar parameters are compatible, within a 3$\sigma$ error, with those determined from spectroscopy in Sect.~\ref{sect_abundances}: $T_{\rm eff} = $24\,500$\pm$1000 K and $\log\,g=$3.45$\pm$0.15. 

The outcome for $f_1$ is displayed in the top panel of Fig.\,\ref{ratios}. It confirms that the dominant mode is a radial one as already ascertained by Aerts\ (\cite{aerts00}). To identify the $\ell$-values of the two observed low-amplitude modes, we then considered models only in the 3$\sigma$ error box that also fit $f_1$ as a radial mode. The amplitude ratios of $f_2^p$ and $f_3^p$ are plotted in the middle and bottom panels of Fig.\,\ref{ratios}, respectively. We conclude that $f_2^p$ is unambiguously identified as a g-mode pulsation with $\ell = 3$. Moreover, $f_3^p$ corresponds to $\ell =$ 0 or 3, with a preference for $\ell = 0$.
 
It is noteworthy that a low frequency such as 0.30818\,d$^{-1}$, which corresponds to a high-order g mode, is not yet a common feature observed for most of the $\beta\,$Cephei stars. However, it is certainly not exceptional. For several of the $\beta\,$Cephei stars that have indeed been the subject of intensive observing campaigns, low frequencies have been found, e.g., in 19\,Mon (Balona et al.\ \cite{balona}: 0.17019\,d$^{-1}$), $\nu\,$Eri (Handler et al.\ \cite{handler04}: 0.43218\,d$^{-1}$), and 12\,Lac (Handler et al.\ \cite{handler06}: 0.35529\,d$^{-1}$). 
\section{Spectroscopy}

\subsection{Data description}

In addition to the photometric datasets, we have at our disposal 262 useful (S/N $>$ 80) high-resolution spectra gathered with three instruments. HD\,180642 was observed during 15 nights (25 June--4 July; 16--20 July 2007) with FEROS at the 2.2m ESO/MPI telescope at La Silla and during 12 nights (13--24 June 2007) with SOPHIE at the 1.93m telescope at OHP in the framework of the CoRoT ground-based Large Programme (Uytterhoeven \& Poretti\ \cite{uytterhoeven_poretti}; Uytterhoeven et al.\ \cite{uytterhoeven08}). Additionally, 22 Aur\'elie spectra at the 1.52m telescope (OHP) and 11 FEROS spectra were already taken in 2005 and 2006 (14, 17, 19, and 22 June 2005 at OHP; 25--26 May 2006 at ESO). An overview and logbook of the spectroscopic observations are given in Table~\ref{logbook2}. 

We carefully calculated the integration times that we selected to ensure a S/N ratio of about 200 near the wavelength 4550 $\AA$, which is the position of a prominent \ion{Si}{iii} triplet well suited to our line-profile variability study (Aerts \& De Cat\ \cite{aerts_decat}). These silicon lines are indeed sufficiently strong without being much affected by blending. Moreover, they are dominated by temperature broadening, such that the intrinsic profile can be modelled by a Gaussian. This simplifies the modeling of the line-profile variations for mode identification purposes (De Ridder et al.\ \cite{deridder}). 

We reduced the FEROS spectra using an improved version of the standard FEROS pipeline, written in MIDAS, developed by Rainer\ (\cite{rainer}). The main improvements of this pipeline concern the blaze and flat-field correction of the spectra, by using an accurate definition of the blaze function extracted from a well-exposed spectrum of a hot star. The SOPHIE spectra were extracted and automatically reduced in real-time by a reduction package adapted from HARPS. Since there is no pipeline reduction available for the Aur\'elie spectrograph, we used standard reduction procedures with IRAF. After a correction to the heliocentric frame, the spectra were manually normalised using a cubic spline fit.  

\subsection{Abundance analysis}\label{sect_abundances}
The non-local thermodynamic equilibrium (NLTE) abundances of helium and the dominant metals have been calculated using the latest versions of the line-formation codes DETAIL/SURFACE (Butler \& Giddings\ \cite{butler_giddings}; Giddings\ \cite{giddings}) and plane-parallel, fully line-blanketed atmospheric models with a solar helium abundance (Kurucz\ \cite{kurucz93}). Our analysis is based on a mean FEROS spectrum created by coadding 11 individual exposures obtained in May 2006. All exposures were transferred to the laboratory rest frame prior to this operation. Curve-of-growth techniques were used to determine the abundances using the equivalent widths of a set of unblended lines. The reader is referred to Morel et al.\ (\cite{morel06}) for complete details of the methodology used to derive the elemental abundances. 

\subsubsection{Atmospheric parameters}\label{subsect_parameters}
\begin{figure}
\centering
\includegraphics[width=8.5cm]{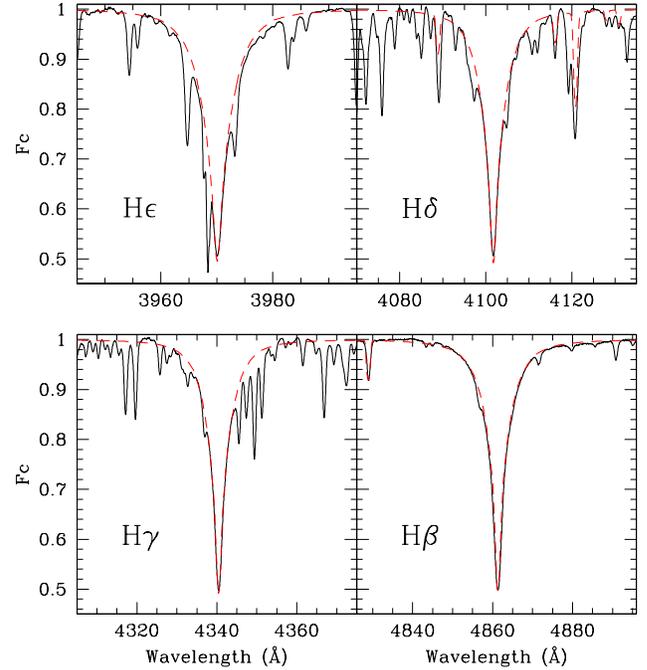}
\caption{Comparison between the synthetic ({\em dashed line}; $T_{\rm eff}$=24\,500 K and $\log g$=3.45 [cgs]) and observed ({\em solid line}) Balmer line profiles. The synthetic spectra have been convolved with a rotational profile with $v$=44 km s$^{-1}$. No attempts have been made to fit the other spectral lines.}
\label{fig_balmer}
\end{figure}

A standard, iterative scheme was used to self-consistently derive the atmospheric parameters solely by spectroscopic means: $T_{\rm eff}$ was determined from the \ion{Si}{ii/iii/iv} ionisation balance, $\log g$ from fitting the collisionally-broadened wings of the Balmer lines, and the microturbulent velocity, $\xi$, by requiring that the abundances corresponding to the \ion{O}{ii} features were independent of the line strength. We obtain: $T_{\rm eff}$=24\,500$\pm$1000 K, $\log g$=3.45$\pm$0.15 dex and $\xi$=12$\pm$3 km s$^{-1}$. As can be seen in Fig.\ref{fig_balmer}, satisfactory fits are obtained for all Balmer lines considered. Within the uncertainties, identical values are obtained when analysing these data with the unified code FASTWIND (Puls et al.\ \cite{puls}) and (semi-)automatic line-profile fitting techniques (Lefever et al.\ \cite{lefever09}): $T_{\rm eff}$=24\,000$\pm$1000 K, $\log g$=3.4$\pm$0.1 dex, and $\xi$=13$\pm$2 km s$^{-1}$. Support for our $T_{\rm eff}$ estimate is provided by the ionisation balance of neon, which suggests an identical value: $T_{\rm eff}$=24\,000$\pm$1000 K (Morel \& Butler\ \cite{morel_butler}). Only a single \ion{C}{iii}, \ion{N}{iii}, or \ion{S}{iii} line could be measured, but good agreement with the abundances yielded by the far more numerous transitions of lower ionisation ions (namely \ion{C}{ii}, \ion{N}{ii}, and \ion{S}{ii}) is also found in each case. The fulfillment of ionisation balance for all species under investigation indicates that the effective temperature of HD\,180642 is well constrained. 

Given the large amplitude of the radial mode, we investigated the amplitude of the variation in atmospheric parameters over the pulsation cycle by analysing separately the average of 20 FEROS exposures corresponding to minimum and maximum EWs of the \ion{Si}{iii} lines, and obtained $\Delta T_{\rm eff}$$\sim$1000 K and $\Delta \log g$$\sim$0.1 dex. The same result was already obtained for the $\beta$~Cephei star $\xi^1$~CMa, which resembles HD\,180642 from the viewpoint of the dominant mode (Morel et al.\ \cite{morel06}). Finally, we infer a total amount of line broadening arising from pulsations and rotation, $v_T$=44 km s$^{-1}$, by comparing the profiles of a set of isolated \ion{O}{ii} lines with a grid of rotationally-broadened synthetic spectra. The stellar parameters suggest that HD\,180642 is an object close to the end of the core-hydrogen burning phase. We refer to a forthcoming paper on modeling (Thoul et al.,\ in prep.) for an in-depth discussion of its evolutionary status.

\subsubsection{Elemental abundances}\label{subsect_abundances}
The abundances computed for $T_{\rm eff}$=24\,500 K, $\log g$=3.45 dex, and $\xi$=12 km s$^{-1}$ are given in Table~\ref{tab_abundances}, and compared with the standard solar mixture of Grevesse \& Sauval\ (\cite{grevesse_sauval}) and values derived from time-dependent, three-dimensional hydrodynamical models (Asplund et al.\ \cite{asplund}). The quoted uncertainties take into account both the line-to-line scatter and the errors arising from the uncertainties in the atmospheric parameters. We note that the strong, numerous \ion{N}{ii} features can also be used to constrain the microturbulence, and suggest a significantly lower value than the \ion{O}{ii} lines: $\xi$=7 km s$^{-1}$ (see, e.g., Trundle  et al.\ \cite{trundle} for a general discussion of this problem). Adopting this microturbulent velocity would lead to a slight ($\Delta\log \epsilon$ $\lesssim$ 0.2 dex) upward revision of the abundances and metallicity (see Table~\ref{tab_micro}) bringing them into closer agreement with the probably more reliable solar abundances of Asplund et al. (\cite{asplund}).

\begin{table}
\begin{center}
\caption{Mean NLTE abundances (on the scale in which $\log \epsilon$[H]=12) and total 1-$\sigma$ uncertainties (in brackets), in units of the last digits.}
\label{tab_abundances}
\begin{tabular}{lcccc} \hline\hline
                         & HD\,180642   & B stars      & Sun 1-D    & Sun 3-D\\\hline
He/H(4)                  & 0.088(18)    & $\sim$0.089  & 0.085(1)   & 0.085(2)\\
$\log \epsilon$(C)\ (9)  & 8.21(10)     & $\sim$8.20   & 8.52(6)    & 8.39(5)\\ 
$\log \epsilon$(N) (21)  & 8.00(19)     & $\sim$7.79   & 7.92(6)    & 7.78(6)\\ 
$\log \epsilon$(O) (25)  & 8.53(14)     & $\sim$8.58   & 8.83(6)    & 8.66(5)\\ 
$\log \epsilon$(Ne) (6)  & 7.86(16)$^a$ & $\sim$8.07   & 8.08(6)    & 7.84(6)\\ 
$\log \epsilon$(Mg) (1)  & 7.34(20)     & $\sim$7.48   & 7.58(5)    & 7.53(9)\\ 
$\log \epsilon$(Al) (3)  & 6.22(15)     & $\sim$6.08   & 6.47(7)    & 6.37(6)\\ 
$\log \epsilon$(Si) (7)  & 7.19(19)     & $\sim$7.20   & 7.55(5)    & 7.51(4)\\ 
$\log \epsilon$(S) (4)   & 7.10(34)     & $\sim$7.19   & 7.33(11)   & 7.14(5)\\ 
$\log \epsilon$(Fe) (21) & 7.34(21)     & $\sim$7.36   & 7.50(5)    & 7.45(5)\\
\hline
${\rm [N/C]}$            &  --0.21(22)  & $\sim$--0.41 & --0.60(9)  & --0.61(8)\\
${\rm [N/O]}$            &  --0.53(24)  & $\sim$--0.79 & --0.91(9)  & --0.88(8)\\
\hline
$Z$                      & 0.0099(16)   & $\sim$0.0108 & 0.0172(12) & 0.0124(7)\\
\hline
\end{tabular}
\end{center}
\tiny{The number of spectral lines used is given in brackets in the first column. For comparison purposes, we provide typical values found for early B dwarfs in the solar neighbourhood (Morel\,\cite{morel09}, and references therein; B stars), the standard solar composition of Grevesse \& Sauval\ (\cite{grevesse_sauval}; Sun 1-D), and updated values derived from three-dimensional hydrodynamical models (Asplund et al.\ \cite{asplund}; Sun 3-D). We define [N/C] and [N/O] as $\log$[$\epsilon$(N)/$\epsilon$(C)] and $\log$[$\epsilon$(N)/$\epsilon$(O)], respectively. The metallicity, $Z$, is given in the last row. To compute this quantity, we assumed the abundances of Grevesse \& Sauval\ (\cite{grevesse_sauval}) for the trace elements not under study.\\
$^a$ A negligible difference amounting to 0.01 dex is obtained when using the $T_{\rm eff}$ derived from the Ne ionisation balance (Morel \& Butler\ \cite{morel_butler}).}\\
\end{table}

\begin{table}
\caption{Abundance and metallicity differences when adopting the microturbulent velocity yielded by the \ion{N}{ii} lines ($\xi$=7 km s$^{-1}$), instead of the adopted value estimated from the \ion{O}{ii} features ($\xi$=12 km s$^{-1}$).}
\label{tab_micro}
\begin{center}
\begin{tabular}{lcc} \hline\hline
$\Delta \xi$              & from 12 to 7 km s$^{-1}$ \\\hline
$\Delta$He/H              & +0.025\\
$\Delta\log \epsilon$(C)  & +0.04 \\ 
$\Delta\log \epsilon$(N)  & +0.13 \\ 
$\Delta\log \epsilon$(O)  & +0.16 \\ 
$\Delta\log \epsilon$(Ne) & +0.03 \\ 
$\Delta\log \epsilon$(Mg) & +0.15 \\ 
$\Delta\log \epsilon$(Al) & +0.11 \\ 
$\Delta\log \epsilon$(Si) & +0.10 \\ 
$\Delta\log \epsilon$(S)  & +0.02 \\ 
$\Delta\log \epsilon$(Fe) & +0.06 \\ 
\hline
${\rm [N/C]}$             & +0.09\\
${\rm [N/O]}$             & --0.03\\
\hline
$Z$                       & +0.0027 \\
\hline
\end{tabular}
\end{center}
\end{table}

The logarithmic ratios of the CNO abundances [N/C] to [N/O] point towards a mild nitrogen overabundance, as unexpectedly observed in other slowly-rotating $\beta$ Cephei stars (Morel et al.\ \cite{morel06}). For HD\,180642, however, this result cannot be firmly established owing to the relatively large error bars and the lack of boron data, which could in principle be used to corroborate (or refute) the existence of such an N excess (see Morel et al.\ \cite{morel08}). The abundances of the other metals are fully compatible with literature values for nearby main-sequence B stars (e.g., Daflon \& Cunha\ \cite{daflon_cunha}; Gummersbach et al.\ \cite{gummersbach}; Kilian-Montenbruck et al.\ \cite{kilian}). We note that the global metallicity, $Z=$0.0099$\pm$0.0016, is very well constrained, because the species analysed account for up to 97\% of the total metal content. 

\subsection{Frequency analysis}\label{spectro_FA}
 
To increase the S/N of the spectra and improve the detection of the line-profile variations, we combined several line profiles as follows. First, cross-correlated profiles were computed using the leastsquares deconvolution (LSD) method (Donati et al.\ \cite{donati}) by means of a mask consisting of lines from the VALD database (Piskunov et al.\ \cite{piskunov}; Ryabchikova et al.\ \cite{ryabchikova}; Kupka et al.\ \cite{kupka}) for $T_{\rm eff}=24~000$ K and $\log g = 3.50$. All elements were taken into account, except He and H. This resulted in combining about 570 lines for the FEROS and SOPHIE spectra and only 78 lines for the Aur\'elie spectra, leading to average S/N ratios of 600, 400, and 90, respectively. 

Second, to avoid the effects of the different behaviour of the individual lines during the pulsation cycle, which was ignored in the first method, we computed mean spectra of sufficiently strong and unblended lines originating from the same element and ionisation state. The averaged spectrum was computed for \ion{He}{i} lines, \ion{N}{ii} lines, \ion{O}{ii} lines, and \ion{Si}{iii} lines. The use of LSD profiles or different elements did not reveal additional periodicities. In what follows, we describe only our frequency analysis of the combination of the two deepest silicon lines of the \ion{Si}{iii} triplet around \mbox{4567\,$\AA$}, of an average S/N ratio of 350.

To perform our frequency analysis, we used the software package FAMIAS\footnote{FAMIAS has been developed in the framework of the FP6 European Coordination Action HELAS -- http://www.helas-eu.org/} (Zima\ \cite{zima08}). We first examined the first three moments \mbox{$<v^1>$}, \mbox{$<v^2>$} and \mbox{$<v^3>$} (see Aerts et al.\ \cite{aerts92} for a definition) of the combined silicon lines. Since the line profiles move significantly because of pulsation, the integration limits for computing the moments were dynamically chosen by sigma clipping to avoid the noisy continuum (see Zima\ \cite{zima08}). 

Harmonics up to the 5{\it th} of the dominant frequency are of significant amplitude in \mbox{$<v^1>$}. After removing them, we do find four additional significant frequencies in the residuals of \mbox{$<v^1>$}. All frequencies retained, except the last one, have an amplitude larger than the 4 S/N-level, computed in the residuals, in a \mbox{2\,d$^{-1}$} interval centered on the frequency of interest. The reason for retaining the last peak is that it is a one-day-alias of a combination frequency. In our frequency solution, we use the value of the combination frequency. The detected frequency values and their corresponding amplitudes and phases of the leastsquares sine fits to \mbox{$<v^1>$} are listed in Table\,\ref{a_p}. These frequencies were also found in the CoRoT light curve (Degroote et al.\ \cite{degroote}), albeit not with exactly the same values. We already discussed a probable explanation of these differences in Sect.\,\ref{freqsection}. When comparing with our ground-based photometry, we note that the low-frequency mode is also found, while the photometric frequency at $\sim 7.36$\,d$^{-1}$ is not detected. Moreover, we detect additional periodicities in the spectroscopic dataset. A frequency analysis in \mbox{$<v^2>$} and \mbox{$<v^3>$} did not detect additional periodicities.

\begin{figure}
\vspace{1.2cm} 
\hspace{0.7cm}\includegraphics[bb=60 50 450 500, width=13cm]{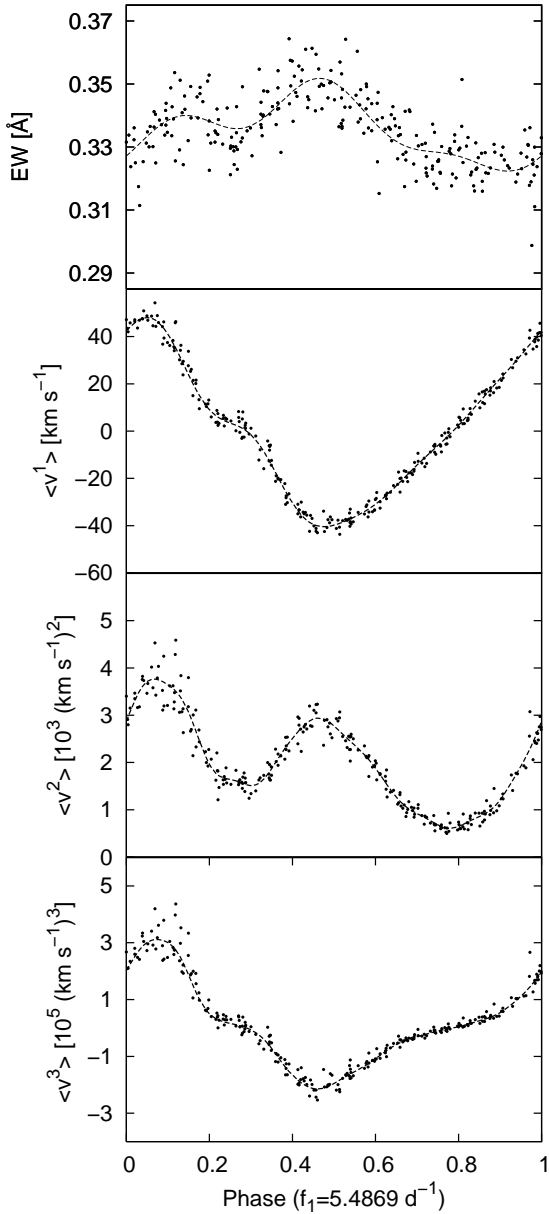}
\caption{Phase diagram of the equivalent width of the \ion{Si}{iii}\,4552\,$\AA$ line, and of the first three velocity moments, for $f_1 = 5.4869\ \rm{d}^{-1}$. The dashed lines represent our best model for the moment variations (see Sect.~\ref{Spectro_modeID} for more explanations).}  
\label{v1v2v3}  
\end{figure} 

Phase diagrams of \mbox{$<v^1>$}, \mbox{$<v^2>$}, and \mbox{$<v^3>$} for the dominant mode are shown in Fig\,.\ref{v1v2v3}. Several characteristics are noticeable in \mbox{$<v^1>$}. As already indicated by the photometry, the dominant mode does not behave sinusoidally at all and \mbox{$<v^1>$} has a peak-to-peak amplitude of about 90 km\,s$^{-1}$. This makes HD\,180642 the $\beta\,$Cephei star with the third largest peak-to-peak radial-velocity amplitude, BW\,Vul having $\sim$~180\,km\,s$^{-1}$ (Crowe \& Gillet\ \cite{crowe_gillet}; Aerts et al.\ \cite{aerts95}; Mathias et al.\ \cite{mathias98}) and $\sigma\,$Sco having $\sim$~110\,km\,s$^{-1}$ (Mathias et al.\ \cite{mathias91}). More unusual is the presence of a so-called stillstand, i.e., a plateau at half the descending branch of the radial velocity curve. In the ground-based photometry, this stillstand phenomenon is well observed at certain epochs but is not clearly visible in the phased light curves of the merged datasets (Fig.\,\ref{fases1}). Such a stillstand is also encountered in BW\,Vul and $\sigma$~Sco. These common characteristics of the three targets suggest a behaviour of the atmospheric pulsation of HD\,180642 similar to that of BW\,Vul and $\sigma$\,Sco. For the latter stars, the picture requires the passage of two shocks per pulsation period, the stillstand corresponding to a relaxation phase (Mathias et al.\,\cite{mathias98}).

A first indicator of the shock propagation is the evolution in the line profile, which exhibits a doubling phenomenon when the front shock enters the line forming region, following the Schwarzschild mechanism (Schwarzschild\ \cite{schwarzschild}). The second indicator is a phase lag between line forming regions, due to the finite shock front velocity, which is present in progressive waves. For HD\,180642, no line-doubling behaviour is seen in the profiles, but it could be hidden by the presence of the low-amplitude modes. No phase lag is also observed between variations associated with different ions. We therefore conclude that the presence of shocks in HD\,180642 is not established, nor can be rejected. The observation of the star during a few consecutive cycles would allow us to prevent the beat phenomenon with the non-radial modes and identify the same behaviour as observed for BW\,Vul and $\sigma$\,Sco.

The behaviour of \mbox{$<v^2>$} (see middle panel of Fig.\,\ref{v1v2v3}) also deserves some comments. It clearly cannot be described by a double sine, as linear theory predicts for radial pulsators (Aerts et al.\ \cite{aerts92}). In addition, the maxima (minima) at phases 0.05 and 0.45 (at phases 0.25 and 0.80) are different. This difference cannot be attributed to a non-linear velocity at the stellar surface only but it reflects that temperature variations also take place. By taking into account temperature effects, \mbox{$<v^2>$} can be satisfactorily modelled, as shown in Sect.~\ref{Spectro_modeID}. This is because the second moment depends on the thermal width of the local profile, which varies with temperature. Similar behaviour, but less prominent, was already observed for $\delta$~Ceti (Aerts et al.\ \cite{aerts92}) and $\xi^1$~CMa (Saesen et al.\ \cite{saesen}).

\begin{table}
\caption{Frequencies, amplitudes, and phases of the leastsquares sine fits to the observed first moment. }
\begin{center}
\begin{tabular}{cccc}
\hline
\hline\\[-7pt]
ID & Frequency & Amplitude & Phase \\[2pt]
   & (d$^{-1}$) & (km\,s$^{-1}$) & (rad)\\[2pt]                
\hline\\[-7pt]
$f_1$   & 5.4869(1) & 38.8(2) & 3.139(6) \\
$2 f_1$ & 10.9738(1) & 3.0(2) & 5.45(8) \\
$3 f_1$ & 16.4607(1) & 4.7(2) & 3.30(5) \\
$4 f_1$ & 21.9476(1) & 3.6(2) & 5.30(7) \\
$5 f_1$ & 27.4345(2) & 1.7(2) & 0.8(1)  \\
$f_2^s$ & 8.4079(2) & 1.5(2) & 4.2(1) \\
$f_3^s$ & 0.3046(2) & 1.1(2) & 2.0(2) \\
$f_4^s$ & 7.1037(3) & 1.0(2) & 2.7(2) \\
$f_1+f_2^s$ & 13.8948(3) & 0.8(2) & 6.3(3) \\[5pt]
\hline
\end{tabular}
\end{center}
\label{a_p}
\tiny{The error estimates in units of the last digit are given in brackets. The adopted reference epoch for $\phi=0.0$ corresponds to HJD 2447963.2564.}
\end{table}

\begin{figure*}
\centering
\vspace{-2cm}
\hspace{0.2cm}
\includegraphics[bb=370 50 790 600,width=6.85cm]{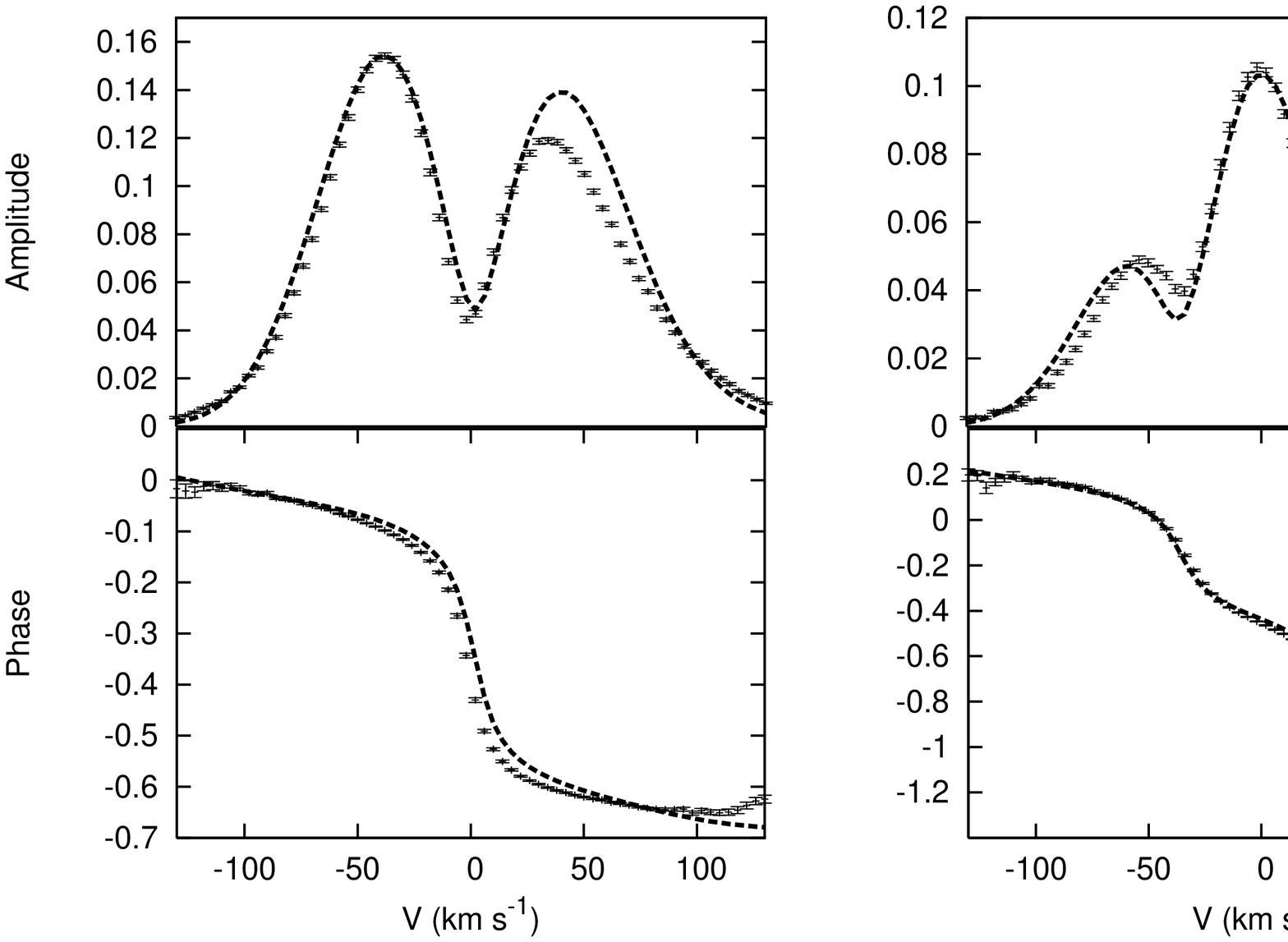}\\[-10pt]
\vspace{-1.5cm}
\includegraphics[bb=370 50 790 600,width=6.85cm]{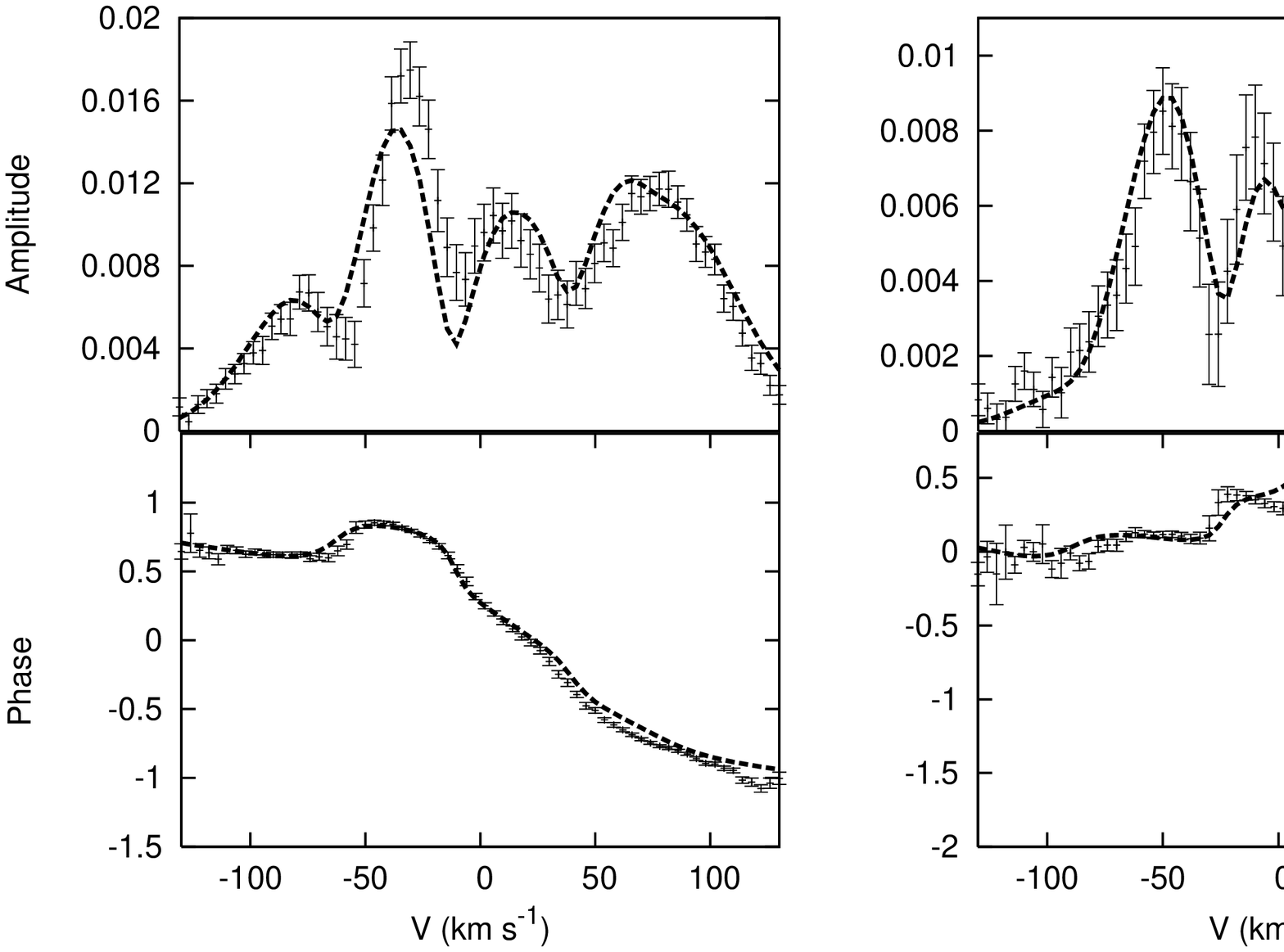}
\caption{Amplitude and phase distributions (points with error bars) across the combination of the two deepest silicon lines of the \ion{Si}{iii} triplet around \mbox{4567\,$\AA$}, for the dominant radial mode. From left to right and top to bottom, we have the frequencies $f_1=5.4869$ d$^{-1}$, 2$f_1$, 3$f_1$, 4$f_1$, 5$f_1$, and 6$f_1$. The bestfit model is represented by dashed lines. The amplitudes are expressed in units of continuum and the phases in $\pi$ radians.}
\label{amp_phase_main_mode}
\end{figure*}

\begin{figure*}
\centering
\vspace{-2cm}
\resizebox{0.75\linewidth}{!}{\rotatebox{-90}{\includegraphics{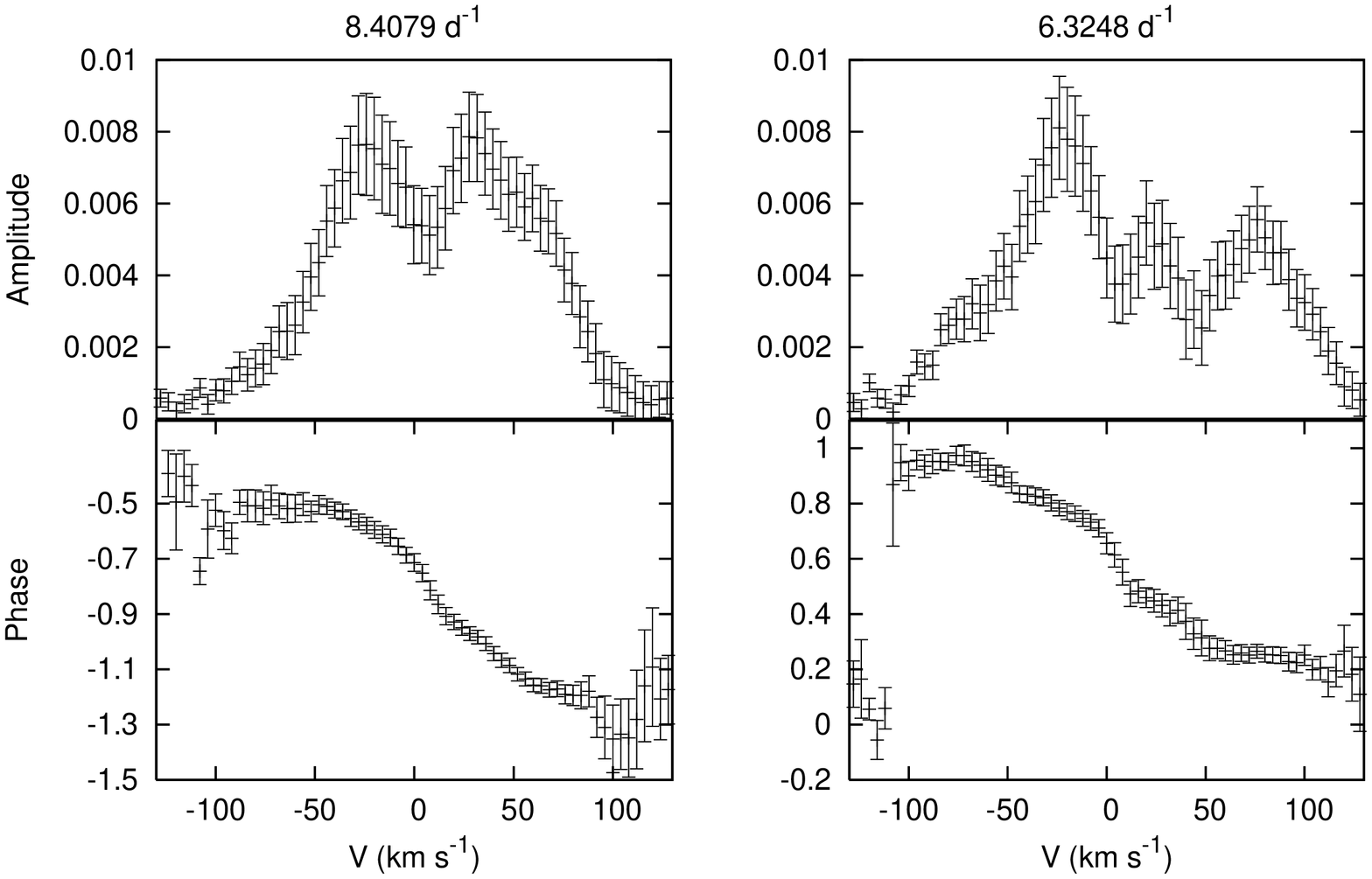}}}\\
\vspace{-1.5cm}
\resizebox{0.75\linewidth}{!}{\rotatebox{-90}{\includegraphics{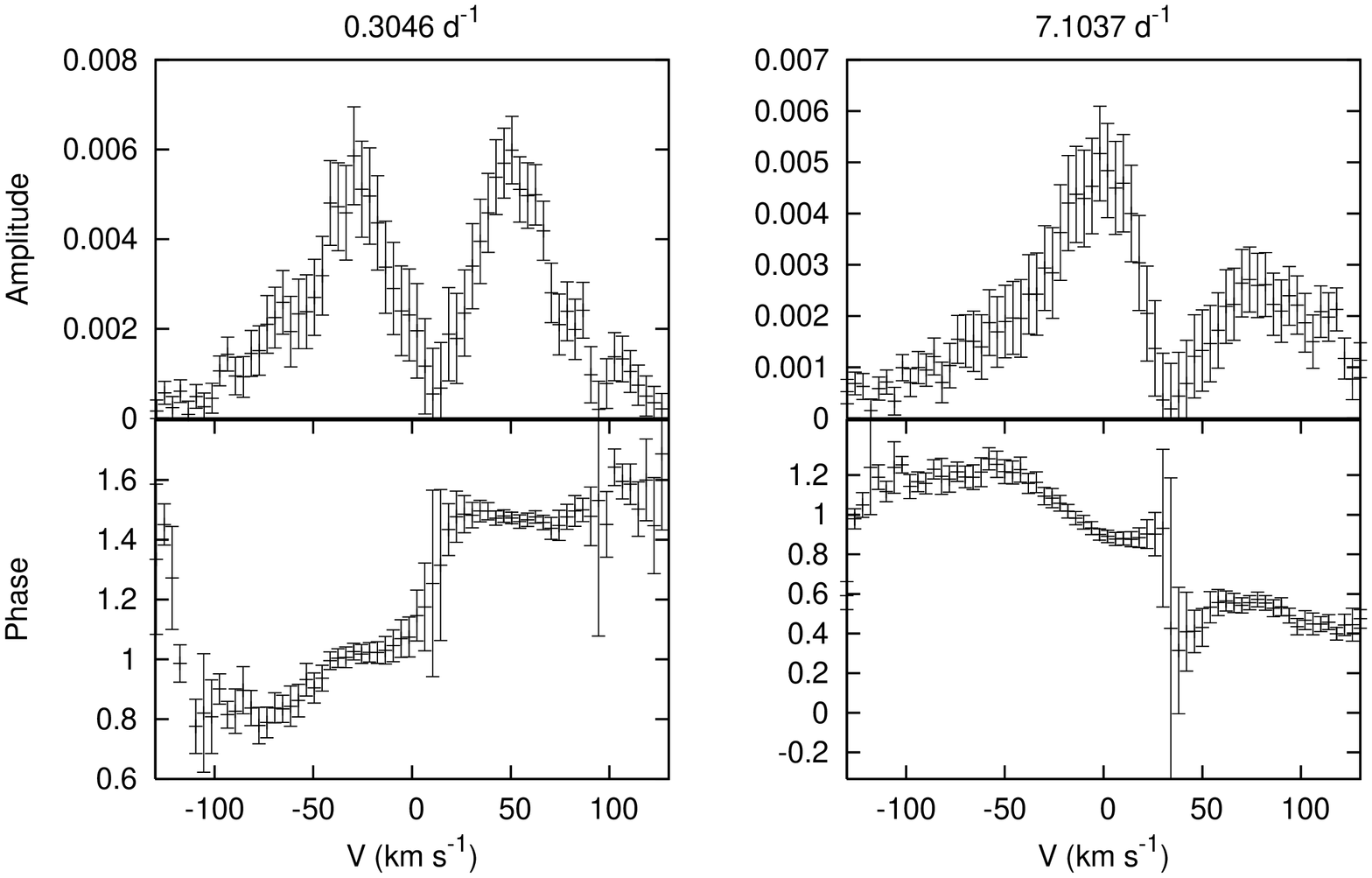}}}
\caption{Amplitude and phase distributions (points with error bars) across the combination of the two deepest silicon lines of the \ion{Si}{iii} triplet around \mbox{4567\,$\AA$}, for the frequencies 8.4079\,d$^{-1}$, 6.3248\,d$^{-1}$, 0.3046\,d$^{-1}$ and 7.1037\,d$^{-1}$. The amplitudes are expressed in units of continuum and the phases in $\pi$ radians.}
\label{amp_phase}
\end{figure*}

The temperature variations caused by the large-amplitude radial mode are also revealed in the equivalent width of the spectral lines. As shown in Fig.\,\ref{v1v2v3}, the equivalent width of the \ion{Si}{iii}\,4552\,$\AA$ line varies with the dominant frequency and two significant harmonics. We point out that this strong EW variability is not common among pulsating B-type stars. We refer to both De Ridder et al.\ (\cite{deridder}) and De Cat\ (\cite{decat02}) for typical examples of $\beta$~Cephei stars and slowly pulsating B stars, respectively. However, this behaviour is certainly not unique, and is observed in several stars with a high-amplitude mode. For $\beta$~Cephei stars, it is also the case for BW~Vul (Mathias et al.\ \cite{mathias98}), $\sigma$~Sco (Mathias et al.\ \cite{mathias91}), $\delta$~Ceti (Aerts et al.\ \cite{aerts92}), $\nu$~Eridani (Aerts et al.\ \cite{aerts04}), and the prototype $\beta$~Cephei itself (Telting et al.\ \cite{telting}, Catanzaro \& Leone\ \cite{catanzaro_leone}). All these targets are monoperiodic radial pulsators or multiperiodic pulsators with a highly dominant radial mode, as is the case for HD\,180642. A different case concerns the $\beta$~Cephei star 12 Lac (Desmet et al.\ \cite{desmet}). This latter object, which also is of relatively high amplitude, shows an equivalent width that varies with its two dominant modes but the highest amplitude mode is identified as an $(\ell,m)=(1,1)$. 

A frequency search was carried out for the spectra by means of a two-dimensional Fourier analysis available in FAMIAS. Once the frequencies are selected, a non-linear multi-periodic leastsquares fit of a sum of sinusoidals is computed with the Levenberg-Marquardt algorithm. This fitting is applied for every bin of the spectrum separately according to the formula $Z+\sum_i A_i \sin\bigl[2\pi (f_i t+\phi_i)\bigr],$ where $ Z$ is the zeropoint, and $ A_i$, $ f_i$, and $ \phi_i$ are the amplitude, frequency, and phase of the $i$-th frequency, respectively. The amplitude and phase distributions across the combined silicon lines are shown, in Fig.\,\ref{amp_phase_main_mode}, for the dominant radial mode and, in Fig.\,\ref{amp_phase}, for the other frequencies detected. 

We emphasize that the 2D frequency analysis detected one additional frequency, which was also observed in the CoRoT photometry ($f_5^s$ = 6.3248\,d$^{-1}$), compared to the 1D analysis of the moments, which are integrated quantities. Our Fourier analysis detected a one-day-alias of $f_5^s$ and we needed the CoRoT data to pinpoint the true periodicity. Additional frequencies were also discovered by a 2D frequency search in the case of several other $\beta$~Cephei stars (see Telting et al.\ \cite{telting} for $\beta$~Cephei; Schrijvers et al.\ \cite{schrijvers} for $\epsilon$~Centauri; Briquet et al.\ \cite{briquet05} for $\theta$~Ophiuchi), pointing out the necessity of performing this kind of analysis, in addition to a frequency search in the radial velocity or higher-order moments.

\subsection{Mode identification}\label{Spectro_modeID}

The moment method (Briquet \& Aerts\ \cite{briquet_aerts}) and the Fourier parameter fit (FPF) method (Zima\ \cite{zima06}) are spectroscopic mode identification techniques appropriate to identifying both the degree $\ell$ and the azimuthal order $m$ of the oscillations of main-sequence pulsators hotter than the Sun. Several successful applications, in particular to $\beta$~Cephei stars, are available in the literature (e.g. Briquet et al.\ \cite{briquet05} for $\theta$~Oph, Mazumdar et al.\ \cite{mazumdar} for $\beta$~CMa, Desmet et al.\ \cite{desmet} for 12 Lac).   

As explained and illustrated in Zima\ (\cite{zima06}), the FPF technique, based on mono-mode line profiles, cannot be applied in case of a large pulsation velocity relative to the projected rotational velocity, i.e., if the radial velocity amplitude is above 0.2 v\,$\sin i$, as for HD\,180642. Instead of using amplitude and phase distributions for mono-mode line profiles, we thus considered the same quantities but computed them for a huge grid of multiperiodic line profile time series. We fixed the dominant mode as radial and its velocity field at the stellar surface was modelled as the superposition of the velocity fields of five linear radial modes.

To model the local temperature and surface gravity variations, we adopted the following empirical approach. At each time of observation, the equivalent width (EW) and the width $\sigma$ of an intrinsic Gaussian profile were chosen to match the observed EW variations and second moment variations. We point out that this model assumes that the temperature and gravity variations are due solely to the dominant mode, which is a reasonable assumption in view of the very low amplitudes of the other modes. The theoretical moment variations in this simple model are shown in Fig.\,\ref{v1v2v3}. The amplitude and phase distributions are displayed in Fig.\,\ref{amp_phase_main_mode}. Our modeling of the non-linear radial mode reproduces the observed behaviour quite well, especially for the first four harmonics. 

To identify the wavenumbers ($\ell,m$) of the low-amplitude modes, we added their pulsational velocity field to the one of the radial mode modelled as described above, and, using a $\chi^2$ value as in Zima\ (\cite{zima06}), we searched for the solutions that fit the observed amplitude and phase distributions computed from multi-mode line profiles best. Unfortunately, no good fit could be achieved for any of the modes. Next, we used the moment method for multiperiodic stars (version of Briquet \& Aerts\ \cite{briquet_aerts}) also adapted to adequately model the dominant radial mode, but, again, the outcome was inconclusive. 

\begin{table}
\caption{The bestfit solutions of the spectroscopic mode identification for the mode with frequency $8.4079$ d$^{-1}$ determined by the adapted \mbox{discriminant $\Sigma$}, based on the definition in Aerts\ (\cite{aerts96}). 
 }
\label{discri}
\begin{center}
\begin{tabular}{ccccccc}
\hline
\hline\\[-7pt]
($\ell,|m|$) & $i$ & v\,$\sin i$ & A$_p$ & v$_{\rm r,max}$ & v$_{\rm t,max}$ & $\Sigma$ \\[2pt]              
\hline\\[-7pt]
(3,2) & 81.5 & 38.5 & 77.35 & 30.4 & 3.0 & 0.93 \\
(3,1) & 30.5 & 39.5 & 25.69 & 11.4 & 1.3 & 4.13 \\
(1,1) & 14.5 & 29.9 & 31.73 & 11.0 & 0.4 & 4.41 \\
(2,1) & 14.0 & 25.2 & 25.14 & 9.7  & 0.7 & 4.42 \\
(2,2) & 47.5 & 38.4 & 21.71 & 8.4  & 0.6 & 4.48 \\
(3,3) & 66.5 & 23.8 & 35.52 & 14.8 & 1.7 & 4.59 \\
(3,0) & 23.0 & 1   &  26.90 & 20.0 & 1.7 & 35.63 \\
(1,0) & 83.5 & 1   &  49.63 & 24.2 & 1.0 & 35.65 \\
(2,0) & 63.0 & 1   &  37.88 & 23.8 & 1.4 & 35.65 \\
(0,0) & 90.0 & 1   &   7.72 & 2.2  & 0.0 & 36.45 \\
(4,0) & 3.0  & 1   & 101.88 & 86.0 & 9.1 & 49.94 \\[5pt]
\hline
\end{tabular}
\end{center}
\tiny{The inclination angle $i$ is expressed in degrees; v\,$\sin i$ is the projected rotational velocity, expressed in km s$^{-1}$; $A_p$ is the amplitude of the radial part of the pulsation velocity, expressed in \mbox{km s$^{-1}$}; and v$_{\rm r,max}$ and v$_{\rm t,max}$ are, respectively, the maximum radial and tangential surface velocity due to the mode, expressed in \mbox{km s$^{-1}$}.}
\end{table}

Therefore, we tried to identify the modes individually, following the version of the moment method of Aerts\ (\cite{aerts96}). Since the width $\sigma$ of the intrinsic profile is not constant in time, in contrast to the assumption in Aerts\ (\cite{aerts96}), we defined a slightly different discriminant, in which the amplitudes involving $\sigma$ are not taken into account. We also omitted the constant term of the second moment because it contains the contribution of all the modes. Therefore, the discriminant used is the one of Aerts\ (\cite{aerts96}) but only with the amplitudes denoted by C, D, F, and G (see Aerts\ \cite{aerts96}) being taken into account. 

We varied the free parameter of the projected rotation velocity, v\,$\sin i$, from 1 to 40 km s$^{-1}$ with a step 0.1 km s$^{-1}$, and the inclination angle of the star, $i$, from 3 to 90$^{\circ}$ with a step 0.5$^{\circ}$. For each combination $(\ell,m,i)$, the amplitude of the mode A$_p$ (given by definition of Briquet \& Aerts\ \cite{briquet_aerts}) was not a free parameter but was determined to fit the observed amplitude of the first moment, within its error. For the low-frequency mode, we fixed $\ell$ to be 3, as derived from the photometry. Finally, the criterion in Sect.\,2.5 in Briquet \& Aerts\ (\cite{briquet_aerts}) excluded $\ell \ge 4$, except ($\ell,m$) = (4,0), for the mode with frequency 8.4079 d$^{-1}$. For the other modes, this criterion limited neither $\ell$ nor $m$.

We obtained a safe mode identification for the mode with frequency 8.4079 d$^{-1}$ only. The discriminant values for the other modes turned out to be too high for us to be confident about their outcome. The discriminant values for the identified mode are given in Table\,\ref{discri}. It does not allow us to determine the sign of $m$. However, prograde and retrograde modes can be distinguished by means of a blue-to-red and red-to-blue respectively descent of the phase in the profile. From Fig.\,\ref{amp_phase}, we deduce that the mode is prograde and denote it with a positive $m$-value. From Table\,\ref{discri}, we conclude that the mode is unambiguously identified as ($\ell,m$) = (3,2). 

\begin{figure} 
\resizebox{0.95\linewidth}{!}{\rotatebox{-90}{\includegraphics{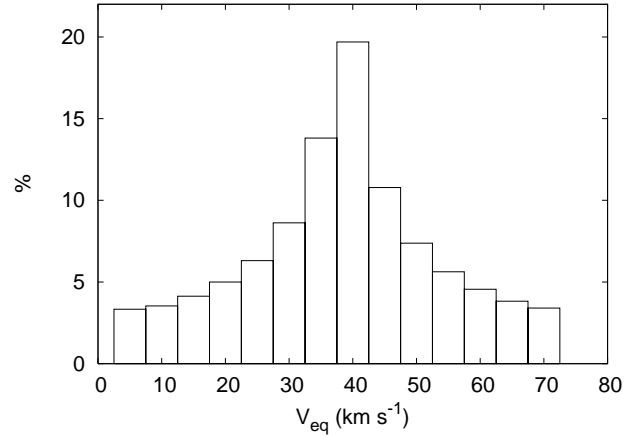}}}
\caption{Histogram for the equatorial rotational velocity of HD\,180642 derived from the moment method outcome, using the discriminant as a weight.}  
\label{histo}  
\end{figure}

By using the discriminant $\Sigma$ as a weight, we constructed histograms for $i$, v\,$\sin i$, and the equatorial rotational velocity v$_{\rm eq}$, as in Mazumdar et al.\ (\cite{mazumdar}). We computed them by considering only solutions with ($\ell,m$) = (3,2). For $i$ and v\,$\sin i$, the distributions are almost flat. However, v$_{\rm eq}$ can be constrained (see Fig.\,\ref{histo}). By calculating a weighted mean and standard deviation, we obtained v$_{\rm eq} =$ 38$\pm$15 km s$^{-1}$. 

\section{Summary}

The $\beta$\,Cephei star HD\,180642 was observed by the CoRoT satellite during a run of 156 days in 2007. The space white-light photometry revealed the rich frequency spectrum of the star (Degroote et al.\,\cite{degroote}). In the present study, we have provided additional information about the target, based on both ground-based multi-colour photometry and high-resolution spectroscopy. We placed our object in the \mbox{(T$_{\rm eff}$, $\log$ g)} diagram. In addition, we derived the chemical abundances of several elements as well as the metallicity of HD\,180642. Finally, we placed constraints on the identification of some modes.        

From the spectroscopic data, we determined $T_{\rm eff} =$ 24\,500$\pm$1000 K and $\log g =$ 3.45$\pm$0.15 dex. In addition, a detailed NLTE abundance analysis showed that the considered abundance values are compatible with values found for B stars in the solar neighbourhood, except for a mild nitrogen excess. This nitrogen overabundance was also discovered for several other $\beta$\,Cephei stars (Morel et al.\,\cite{morel06}, \cite{morel08}). The deduced metallicity for HD\,180642 is \mbox{$Z =$ 0.0099$\pm$0.0016} (or 0.0126$\pm$0.0016 depending on the choice of microturbulence). 

In the photometry, three pulsation frequencies were found to be significant: \mbox{5.48694 d$^{-1}$}, \mbox{0.30818 d$^{-1}$}, and \mbox{7.36673 d$^{-1}$}. The first mode is highly dominant with an amplitude in the U-filter about 15 times larger than that of the other modes. By means of the method of photometric amplitude ratios, we unambiguously identified the degree of the first two modes as $\ell = 0$ and $\ell = 3$. For the frequency 7.36673 d$^{-1}$, two possibilities remained, namely $\ell = 0$ or 3. 

In the radial velocity measurements, the dominant radial mode is highly non-linear, presents a so-called stillstand and has a peak-to-peak amplitude of $\sim$90 km s$^{-1}$. This behaviour is typical of the presence of shocks. However, our dataset did not allow us to prove (or definitely refute) their existence in HD\,180642. 

We attempted mode identification for several low-amplitude modes found in the spectroscopy. In our first trials, we used multiperiodic versions of the moment method and FPF method. Both techniques were adapted and implemented to adequately model the dominant radial mode but we failed for the low-amplitude modes. However, a successful outcome for one of the modes was achieved by means of the discriminant of Aerts\ (\cite{aerts96}), which we adapted slightly to our case. This discriminant identifies the modes independently. In the presence of a highly dominant mode, it is thus more robust than the multiperiodic discriminant of Briquet \& Aerts\ (\cite{briquet_aerts}). We found the frequency $8.4079$ d$^{-1}$ to correspond to a $(\ell,m)=(3,2)$. Finally, we derived a value for the equatorial rotational velocity of v$_{\rm eq}=$ 38$\pm$15 km s$^{-1}$.

All these observational constraints, together with the CoRoT results, will be used to compute stellar models of HD\,180642. This will be presented in a forthcoming paper.

\begin{acknowledgements}
We thank our colleagues from the Institute of Astronomy of Leuven University who contributed to the gathering of the photometric data. The FEROS data were obtained as part of the ESO programme 077.D-0311 (PI: K. Uytterhoeven) as well as as part of the ESO Large Programme 178.D-0361 (PI: E. Poretti). This work was supported by the Hungarian ESA PECS project No 98022, by the Research Council of K.U.Leuven under grant GOA/2008/04 and by the European Helio- and Asteroseismology Network (HELAS), a major international collaboration funded by the European Commission's Sixth Framework Programme. KU acknowledges financial support from a {{\it European Community Marie Curie Intra-European Fellowship}}, contract number MEIF-CT-2006-024476. EP and MR acknowledge financial support from the Italian ASI-ESS project, contract ASI/INAF I/015/07/0, WP 03170. JCS acknowledges support by the ``Consejo Superior de Investigaciones Cient\'{\i}ficas'' by an I3P contract financed by the European Social Fund and from the Spanish ``Plan Nacional del Espacio'' under project ESP2007-65480-C02-01. SMR acknowledges a ``Retorno de Doctores'' contract of the Junta de Andaluc\'{\i}a and IAA for carrying out photometry campaigns for CoRoT targets at Sierra Nevada Observatory. EN acknowledges financial support of the N N203 302635 grant from the MNiSW. MB and FC are Postdoctoral Fellows of the Fund for Scientific Research, Flanders. AM is Postdoctoral Researcher, Fonds de la Recherche Scientifique -- FNRS, Belgium. We also would like to thank K. Butler for making the NLTE line-formation codes DETAIL/SURFACE available to us. 
\end{acknowledgements}

{}


\begin{thebibliography}{}

\bibitem[1992]{aerts92} Aerts, C., De Pauw, M., Waelkens, C.\ 1992, A\&A, 266, 294

\bibitem[1995]{aerts95} Aerts, C., Mathias, P., Van Hoolst, T., De Mey, K., Sterken, C., Gillet, D.\ 1995, A\&A, 301, 781

\bibitem[1996]{aerts96} Aerts, C. 1996, A\&A, 314, 115

\bibitem[2000]{aerts00} Aerts, C.\ 2000, A\&A, 361, 245

\bibitem[2003]{aerts_decat} Aerts, C., De Cat, P.\ 2003, SSRv, 105, 453

\bibitem[2004]{aerts04} Aerts, C., De Cat, P., Handler, G., Heiter, U., Balona, L. A., Krzesinski, J., Mathias, P., Lehmann, H., Ilyin, I., De Ridder, J., et al.\ 2004, MNRAS, 347, 463

\bibitem[2006]{aerts06} Aerts, C., De Cat, P., De Ridder, J., Van Winckel, H., Raskin, G., Davignon, G., Uytterhoeven, K.\ 2006, A\&A, 449, 305

\bibitem[2008]{aerts08} Aerts, C.\ 2008, in Massive Stars as Cosmic Engines, Proceedings of the International Astronomical Union, IAU Symposium, Volume 250, p. 237-244

\bibitem[2005]{asplund} Asplund, M., Grevesse, N., \& Sauval, A.J.\ 2005, in Cosmic abundances as records of stellar evolution and nucleosynthesis, ed. T. G. Barnes III, F. N. Bash, ASP Conf. Ser., 336, 25

\bibitem[2002]{balona}Balona, L.A., James, D.J., Motsoasele, P., Nombexeza, B., Ramnath, A., van Dyk, J.\ 2002, MNRAS, 333, 952

\bibitem[1993]{breger} Breger, M., Stich, J., Garrido, R., Martin, B., Jiang, S. Y., Li, Z. P., Hube, D. P., Ostermann, W., Paparo, M., Scheck, M.\ 1993, A\&A, 271, 482

\bibitem[2003]{briquet_aerts} Briquet, M., Aerts, C.\ 2003, A\&A, 398, 687

\bibitem[2005]{briquet05} Briquet, M., Lefever, K., Uytterhoeven, K., Aerts, C.\ 2005, MNRAS, 362, 619 

\bibitem[2007]{briquet07} Briquet, M., Morel, T., Thoul, A., Scuflaire, R., Miglio, A., Montalb\'an, J., Dupret, M.-A., Aerts, C.\ 2007, MNRAS, 381, 1482

\bibitem[1985]{butler_giddings} Butler, K., \& Giddings, J. R.\ 1985, in Newsletter of Analysis of Astronomical Spectra, No.9 (Univ. London)

\bibitem[2008]{catanzaro_leone} Catanzaro, G., Leone, F.\ 2008, MNRAS, 389, 1414

\bibitem[1989]{crowe_gillet} Crowe, R., Gillet, D.\ 1989, A\&A, 211, 365

\bibitem[2004]{daflon_cunha} Daflon, S., \& Cunha, K.\ 2004, \apj, 617, 1115

\bibitem[2002]{decat02} De Cat, P.\ 2002, In: C. Aerts, T.R. Bedding,
  J. Christensen-Dalsgaard (eds.): Radial and Nonradial Pulsations as Probes of
  Stellar Physics, ASP Conf. Ser., 259, 196

\bibitem[2003]{decat_cuypers} De Cat, P., \& Cuypers, J., in ``Interplay of Periodic, Cyclic and Stochastic Variability in Selected Areas of the H-R Diagram''. Edited by C. Sterken, ASP Conf. Ser. 292. San Francisco: Astronomical Society of the Pacific, 2003, p. 377.
  
\bibitem[2007]{decat} De Cat, P., Briquet, M., Aerts, C., Goossens, K., Saesen, S., Cuypers, J., Yakut, K., Scuflaire, R., Dupret, M.-A., Uytterhoeven, K., et al.\ 2007, A\&A, 463, 243

\bibitem[2009]{degroote} Degroote, P., Briquet, M., Catala, C., et al.\ 2009, A\&A, submitted 

\bibitem[2002]{deridder}De Ridder, J., Dupret, M.-A., Neuforge, C., Aerts, C. 2002\ A\&A, 385, 572

\bibitem[2009]{desmet} Desmet, M., Briquet, M., Thoul, A., Zima, W., De Cat, P., Handler, G., Ilyin, I., Krzesinski, J., Lehmann, H., Masuda, S., Mathias, P., Mkrtichian, D.E, Telting, J., Uytterhoeven, K., Yang, S.L.S., Aerts, C.\ 2009, MNRAS, in press (arXiv0903.5477)

\bibitem[1999]{donati} Donati, J.-F., Semel, M., Carter, B.D., Rees, D. E., \& Collier Cameron, A.\ 1997, MNRAS, 291, 658

\bibitem[2001]{dupret01} Dupret, M.-A.\ 2001, A\&A, 366, 166

\bibitem[2003]{dupret03} Dupret, M.-A., De Ridder, J., De Cat, P., Aerts, C., Scuflaire, R., Noels, A., Thoul, A.\ 2003, A\&A, 398, 677

\bibitem[1993]{dziembowski_pamyatnykh} Dziembowski, W.A., Pamyatnykh, A.A.\ 1993, MNRAS, 262, 204

\bibitem[2008]{dziembowski_pamyatnykh08} Dziembowski, W.A., Pamyatnykh, A.A.\ 2008, MNRAS, 385, 2061

\bibitem[1993]{gautschy_saio} Gautschy, A., Saio, H.\ 1993, MNRAS, 262, 213

\bibitem[1981]{giddings} Giddings, J. R. 1981, Ph.D. Thesis , University of London

\bibitem[1998]{grevesse_sauval} Grevesse, N, \& Sauval, A.J.\ 1998, \ssr, 85, 161 

\bibitem[1998]{gummersbach} Gummersbach, C. A., Kaufer, A., Sch\"{a}fer, D. R.,
Szeifert, T., \& Wolf, B.\ 1998, \aap, 338, 881

\bibitem[2004]{handler04} Handler, G., Shobbrook, R.R., Jerzykiewicz, M., Krisciunas, K., Tshenye, T., Rodr\'iguez, E., Costa, V., Zhou, A.-Y., Medupe, R., Phorah, W.M., et al.\ 2004, MNRAS, 347, 454

\bibitem[2006]{handler06} Handler, G., Jerzykiewicz, M., Rodr\'iguez, E., Uytterhoeven, K., Amado, P.J., Dorokhova, T.N., Dorokhov, N.I., Poretti, E., Sareyan, J.-P., Parrao, L., et al.\ 2006, MNRAS, 365, 327

\bibitem[1996]{iglesias_rogers} Iglesias, C.A., Rogers F.J.\ 1996, ApJ 464, 943

\bibitem[1994]{kilian} Kilian-Montenbruck, J., Gehren, T., \& Nissen,
P.E.\ 1994, \aap, 291, 757

\bibitem[1999]{kupka} Kupka, F., Piskunov, N.E, Ryabchikova, T.A., Stempels, H.C. \& Weiss, W.W.\ 1999, A\&ASS, 138, 119

\bibitem[1993]{kurucz93} Kurucz, R. L. 1993, ATLAS9 Stellar Atmosphere Programs
and 2 km/s grid.~Kurucz CD-ROM No.~13.~ Cambridge, Mass.: Smithsonian
Astrophysical Observatory, 1993, 13

\bibitem[1997]{kuschnig}Kuschnig, R., Weiss, W.W., Gruber, R., Bely, P.Y., Jenkner, H.\ 1997, A\&A, 328, 544

\bibitem[2007]{lefever07} Lefever, K.\ 2007, Ph. D. Thesis, Katholieke Universiteit Leuven, Belgium

\bibitem[2009]{lefever09} Lefever, K., Puls, J., Morel, T., Aerts, C., Decin, L., Briquet, M.\ 2009, A\&A, submitted

\bibitem[1991]{mathias91} Mathias, P., Gillet, D., Crowe, R.\ 1991, A\&A, 252, 245

\bibitem[1998]{mathias98} Mathias, P., Gillet, D., Fokin, A.B., Cambon, T.\ 1998, A\&A, 339, 525

\bibitem[2006]{mazumdar}Mazumdar, A., Briquet, M., Desmet, M., Aerts, C.\ 2006, A\&A, 459, 589

\bibitem[2007]{miglio} Miglio, A., Montalb\'an, J., Dupret, M.-A.\ 2007, MNRAS 375, 21

\bibitem[1999]{montgomery_donoghue} Montgomery, M.H., O'Donoghue, D.\ 1999, DSSN, 13, 28

\bibitem[2006]{morel06} Morel, T., Butler, K., Aerts, C., Neiner, C., \& Briquet, M.\ 2006, \aap, 457, 651 

\bibitem[2008]{morel08} Morel, T., Hubrig, S., \& Briquet, M.\ 2008, \aap, 481, 453

\bibitem[2008]{morel_butler} Morel, T., \& Butler, K.\ 2008, \aap, 487, 307

\bibitem[2009]{morel09} Morel, T. 2009, in Evolution and Pulsation of Massive Stars on the Main Sequence and Close to it, CoAst, in press (arXiv0811.4114)

\bibitem[1999]{pamyatnykh99} Pamyatnykh, A.A.\ 1999, Acta Astr., 49, 119 

\bibitem[2008]{pigulski_pojmanski} Pigulski, A., Pojma\'nski, G.\ 2008, A\&A, 477, 917

\bibitem[1995]{piskunov} Piskunov, N.E., Kupka, F., Ryabchikova, T.A, Weiss, W.W. \& Jeffery, C.S.\ 1995, A\&ASS, 112, 525

\bibitem[2007]{poretti} Poretti, E., Rainer, M., Uytterhoeven, K., Cutispoto, G.,Distefano, E., Romano, P.\ 2007, MmSAI, 78, 624

\bibitem[2005]{puls} Puls, J., Urbaneja, M. A., Venero, R., et al.\ 2005, \aap, 435, 669

\bibitem[2003]{rainer} Rainer, M.\ 2003, 'Analisi spettrofotometriche di stelle da usare come targets per la missione spaziale CoRoT', Laurea Thesis, Universit\'a degli Studi di Milano

\bibitem[1999]{ryabchikova} Ryabchikova, T.A., Piskunov, N.E., Stempels, H.C., Kupka, F. \& Weiss, W.W.\ 1999, Physica Scripta, 83, 162

\bibitem[2006]{saesen} Saesen, S., Briquet, M., Aerts, C.\ 2006, CoAst, 147, 109

\bibitem[2006]{saio} Saio, H., Kuschnig, R., Gautschy, A.\ 2006, ApJ, 650, 1111

\bibitem[1982]{scargle} Scargle, J.D.\ 1982, ApJ 263, 835

\bibitem[2004]{schrijvers} Schrijvers, C., Telting, J.H., Aerts, C.\ 2004, A\&A, 416, 1069 

\bibitem[1952]{schwarzschild} Schwarzschild, M.\ 1952, Transactions of the IAU VIII. Oosterhoff P.Th. (Ed.), Cambridge Univ. Press, Cambridge, p. 811

\bibitem[2008]{scuflaire} Scuflaire, R., Th\'eado, S., Montalb\'an, J., Miglio, A., Bourge, P.-O., Godart, M., Thoul, A., Noels, A.\ 2008, Ap\&SS, 316, 83

\bibitem[2005]{seaton} Seaton, M.J.\ 2005, MNRAS, 362, 1

\bibitem[2005]{stankov_handler}Stankov, A., Handler, G.\ 2005, ApJS, 158, 193

\bibitem[1997]{telting} Telting, J.H., Aerts, C., Mathias, P.\ 1997\ A\&A, 322, 493

\bibitem[2004]{trundle} Trundle, C., Lennon, D. J., Puls, J., \& Dufton, P.L.\ 2004, \aap, 417, 217

\bibitem[1971]{vanicek} Van\'{\i}\v{c}ek, P.\ 1971, Ap\&SS, 12, 10

\bibitem[2008]{uytterhoeven08}
Uytterhoeven, K., Poretti, E., Rainer, M., Mantegazza, L., Zima, W., et al.\ 2008, in 'HelasII international conference: Helioseismology, Asteroseismology and MHD Connections', Journal of Physics: Conference Series, IOP Publishing, 118, 2077

\bibitem[2009]{uytterhoeven09} Uytterhoeven, K.\ 2009, CoAst, 158, in press

\bibitem[2007]{uytterhoeven_poretti} Uytterhoeven, K., Poretti, E., \& the CoRoT SGBOWG, 2007, CoAst, 150, 371

\bibitem[1998]{waelkens}Waelkens, C., Aerts, C., Kestens, E., Grenon, M., Eyer, L.\ 1998, A\&A, 330, 215

\bibitem[2006]{zima06} Zima, W.\ 2006, A\&A, 455, 227

\bibitem[2008]{zima08} Zima, W.\ 2008, CoAst, 157, 387

\end{thebibliography}
\end{document}